# Thermodynamics of mixtures with strongly negative deviations from Raoult's law. XV. Permittivities and refractive indices for 1-alkanol + *n*-hexylamine systems at (293.15-303.15) K. Application of the Kirkwood-Fröhlich model

Fernando Hevia[1], Juan Antonio González[1]*, Ana Cobos[1], Isaías García de la Fuente[1], Cristina Alonso-Tristán[2]

[1]G.E.T.E.F., Departamento de Física Aplicada, Facultad de Ciencias, Universidad de Valladolid. Paseo de Belén, 7, 47011 Valladolid, Spain.

[2]Unidad de Investigación Consolidada UIC-011, JCyL. Departamento de Ingeniería Electromecánica, Escuela Politécnica Superior, Universidad de Burgos. Avda. Cantabria s/n. 09006, Burgos, Spain.

*e-mail: jagl@termo.uva.es; Tel: +34-983-423757

# Abstract

Relative permittivities at 1 MHz, $\varepsilon_r$, and refractive indices at the sodium D-line, $n_D$, are reported at 0.1 MPa and at (293.15-303.15) K for the binary systems 1-alkanol + $n$-hexylamine (HxA). Also, their corresponding excess functions are calculated and correlated. Positive values of the excess permittivities, $\varepsilon_r^E$, are encountered for the methanol system, whereas the remaining mixtures show negative values. This reveals that interactions between unlike molecules contribute positively to $\varepsilon_r^E$. This contribution is dominant for the methanol mixture, while those arising from the breaking of interactions between like molecules are prevalent for the remaining mixtures. At $\phi_1$ (volume fraction) = 0.5, $\varepsilon_r^E$ changes in the order: methanol > 1-propanol > 1-butanol > 1-pentanol < 1-heptanol. Similar variation with the chain length of the 1-alkanol is observed for mixtures such as 1-alkanol + heptane, or + cyclohexylamine, and can be explained in terms of the lower and weaker self-association of longer 1-alkanols. The effect of the replacement of HxA by cyclohexylamine, or by aniline, is also shown. Calculations on molar refractions indicate that dispersive interactions in the systems under study increase with the length of the 1-alkanol. The mixtures are studied by means of the application of the Kirkwood-Fröhlich model, and the Kirkwood correlation factors, including the corresponding excess values, are reported.

# 1. Introduction

Mixtures formed by 1-alkanol and amine are a very interesting class of systems as they show a variety of different behaviours. For example, 1-alkanol + linear primary or secondary amine systems are characterized by strongly negative excess molar enthalpies ($H_m^E$). Thus, at 298.15 K and equimolar composition, $H_m^E$/J·mol$^{-1}$ = –3200 (methanol + *n*-hexylamine (HxA)) [1]; –4581 (methanol + diethylamine) [2]. This has been interpreted as the result of two different opposing effects. In the pure liquid state, both 1-alkanols and amines are self-associated by means of O-H---O and N-H---N bonds, respectively. When the mixing process takes place, such bonds are broken, and this process leads to a positive contribution to $H_m^E$. However, new interactions between unlike molecules are simultaneously created, which contributes negatively to $H_m^E$. Therefore, the large and negative $H_m^E$ values reveal that the new O-H---N bonds created are stronger than the O-H---O and N-H---N bonds. Thus, the values of the enthalpy of the hydrogen bonds between methanol and amine estimated from the application of the ERAS model [3] are: –42.4 kJ·mol$^{-1}$ (*n*-hexylamine) [4]; –45.4 kJ·mol$^{-1}$ (diethylamine) [5]. The value used, within this model, for the enthalpy of the H bonds between alkanol molecules is higher: –25.1 kJ·mol$^{-1}$ [3-5]. As a consequence of the strong interactions between unlike molecules, the systems are highly structured. For example, at 298.15 K and $x_1$ = 0.5, $TS_m^E$ ($=H_m^E - G_m^E$; $G_m^E$ molar excess Gibbs energy) is –3758 J·mol$^{-1}$ for the methanol + diethylamine mixture (value determined using $G_m^E$ = –823 J·mol$^{-1}$ [6]). For comparison, we provide similar results for the 1-propanol + hexane system, $TS_m^E$ = (533 ($=H_m^E$) – 1295 ($=G_m^E$)) = –762 J·mol$^{-1}$ [7, 8]. The existence of strong interactions between unlike molecules in this type of solutions is also supported by large and negative excess molar volumes [4, 9-13] and by solid-liquid equilibria measurements, as the corresponding phase diagrams show that complex formation is an important feature of the systems [14]. Interestingly, the replacement of a linear primary amine by aniline leads to very different $H_m^E$/J·mol$^{-1}$ values: –170 (methanol) [15]; 1020 (1-butanol) [16]. This can be explained in terms of a large contribution to $H_m^E$ from the breaking of the strong dipolar interactions between aniline molecules upon mixing. Note that the upper critical solution temperature of the aniline + heptane system is 343.1 K [17].

We have extended the database of 1-alkanol + amine mixtures reporting excess molar volumes [4, 9-13]; dynamic viscosities [11-13]; vapour-liquid equilibria [18]; permittivities ($\varepsilon_r$) and refractive indices ($n_D$) [11-13, 19]. In addition, these systems have been studied by using different models as DISQUAC or ERAS [4, 5, 9, 10, 12, 20-23]; the formalism of the Kirkwood-Buff integrals [24], or the concentration-concentration structure factor ($S_{CC}(0)$) formalism [25]. As a continuation, we provide now $\varepsilon_r$ and $n_D$ measurements over the temperature range (293.15-303.15) K for the systems 1-alkanol + HxA. In addition, the data are analyzed in terms of the Kirkwood-Fröhlich model [26-29], which is a useful approach to gain insight into the mixture structure and interactions.

## 2. Experimental

*2.1 Materials*

Information about the purity and source of the pure compounds used along the experiments is collected in Table 1. They were used without further purification. Table 2 contains their $\varepsilon_r$ values at 1 MHz, densities ($\rho$) and $n_D$ values at 0.1 MPa and at the working temperatures. These results agree well with literature data.

*2.2 Apparatus and procedure*

Binary mixtures were prepared by mass in small vessels of about 10 cm$^3$ with the aid of an analytical balance Sartorius MSU125p (weighing accuracy 0.01 mg), taking into account the corresponding corrections on buoyancy effects. The standard uncertainty in the final mole fraction is 0.0010. Molar quantities were calculated using the relative atomic mass Table of 2015 issued by the Commission on Isotopic Abundances and Atomic Weights (IUPAC) [30]. In order to minimize the effects of the interaction of the compounds with air components, they were stored with 4 Å molecular sieves (except methanol, because measurements were affected). In addition, the measurement cell (see below) was completely filled with the samples and appropriately closed. Different density measurements of pure compounds, conducted along experiments, showed that this quantity remained unchanged within the experimental uncertainty.

Temperatures were measured with Pt-100 resistances, calibrated according to the ITS-90 scale of temperature, against the triple point of the water and the fusion point of Ga. The standard uncertainty of this quantity is 0.01 K for $\rho$ determinations, and 0.02 K for $\varepsilon_r$ and $n_D$ measurements.

Densities were obtained using a vibrating-tube densimeter and sound analyser Anton Paar DSA 5000, which is automatically thermostated within 0.01 K. The calibration procedure has been described elsewhere [31]. The relative standard uncertainty of the $\rho$ measurements is 0.0012.

A Bellingham+Stanley RFM970 refractometer was used for the $n_D$ measurements. The technique is based on the optical detection of the critical angle at the wavelength of the sodium D line (589.3 nm). The temperature is controlled by Peltier modules and its stability is 0.02 K.

The refractometer has been calibrated using 2,2,4-trimethylpentane and toluene at (293.15-303.15) K, following the recommendations by Marsh [32]. The standard uncertainty of $n_D$ is 0.00008.

The $\varepsilon_r$ measurements were performed with the aid an equipment from Agilent. A 16452A cell, which is a parallel-plate capacitor made of Nickel-plated cobalt (54% Fe, 17% Co, 29% Ni) with a ceramic insulator (alumina, $Al_2O_3$), is filled with a sample volume of $\approx 4.8$ cm$^3$. The cell is connected by a 16048G test lead to a precision impedance analyser 4294A, and immersed in a thermostatic bath LAUDA RE304, with a temperature stability of 0.02 K. Details about the device configuration and calibration are given elsewhere [33]. The relative standard uncertainty of the $\varepsilon_r$ measurements (i.e. the repeatability) is 0.0001. The total relative standard uncertainty of $\varepsilon_r$ was estimated to be 0.003 from the differences between our data and values available in the literature, in the range of temperature (288.15-333.15) K, for the following pure liquids: water, benzene, cyclohexane, hexane, nonane, decane, dimethyl carbonate, diethyl carbonate, methanol, 1-propanol, 1-pentanol, 1-hexanol, 1-heptanol, 1-octanol, 1-nonanol and 1-decanol.

## 3. Results

From the experimental $\varepsilon_r$ values at different temperatures, we can also determine the derivative $\left(\partial \varepsilon_r / \partial T\right)_p$ at 298.15 K as the slope of a linear regression of experimental $\varepsilon_r$ values in the range (293.15 – 303.15) K.

Let us denote by $x_i$ the mole fraction of component $i$. The corresponding volume fraction, $\phi_i$, is given by $\phi_i = x_i V_{mi}^* / \left(x_1 V_{m1}^* + x_2 V_{m2}^*\right)$, where $V_{mi}^*$ stands for the molar volume of component $i$. For an ideal mixture at the same temperature and pressure as the mixture under study, the relative permittivity, $\varepsilon_r^{id}$, the derivative $\left[\left(\partial \varepsilon_r / \partial T\right)_p\right]^{id}$, and the refractive index, $n_D^{id}$, are given by [34, 35]:

$$\varepsilon_r^{id} = \phi_1 \varepsilon_{r1}^* + \phi_2 \varepsilon_{r2}^* \qquad (1)$$

$$n_D^{id} = \left[\phi_1 \left(n_{D1}^*\right)^2 + \phi_2 \left(n_{D2}^*\right)^2\right]^{1/2} \qquad (2)$$

$$\left[\left(\frac{\partial \varepsilon_r}{\partial T}\right)_p\right]^{id} = \left(\frac{\partial \varepsilon_r^{id}}{\partial T}\right)_p \qquad (3)$$

where $\varepsilon_{ri}^{*}$ and $n_{Di}^{*}$ denote the relative permittivity and the refractive index of pure species $i$, and $\left(\partial \varepsilon_{r}^{id}/\partial T\right)_{p}$ is calculated from linear regressions as indicated above. The corresponding excess functions, $F^{E}$, are obtained as

$$F^{E} = F - F^{id} \quad , \quad F = \varepsilon_{r}, n_{D}, \left(\frac{\partial \varepsilon_{r}}{\partial T}\right)_{p} \tag{4}$$

Table 3 lists $\phi_{1}$, $\varepsilon_{r}$ and $\varepsilon_{r}^{E}$ values of 1-alkanol (1) + HxA (2) systems as functions of $x_{1}$, in the temperature range (293.15 – 303.15) K. Table 4 contains the corresponding experimental $x_{1}$, $\phi_{1}$, $n_{D}$ and $n_{D}^{E}$ values. The data of $\left[\left(\partial \varepsilon_{r}/\partial T\right)_{p}\right]^{E} = \left(\partial \varepsilon_{r}^{E}/\partial T\right)_{p}$ are collected in Table S1 (supplementary material).

The $F^{E}$ data were fitted to a Redlich-Kister equation [36] by an unweighted linear least-squares regression:

$$F^{E} = x_{1}(1-x_{1})\sum_{i=0}^{k-1} A_{i}(2x_{1}-1)^{i} \tag{5}$$

The number, $k$, of necessary coefficients for this regression has been determined, for each system and temperature, by applying an F-test of additional term [37] at 99.5% confidence level. Table 5 includes the parameters $A_{i}$ obtained, and the standard deviations $\sigma(F^{E})$, defined by:

$$\sigma(F^{E}) = \left[\frac{1}{N-k}\sum_{j=1}^{N}\left(F_{cal,j}^{E} - F_{exp,j}^{E}\right)^{2}\right]^{1/2} \tag{6}$$

where the index $j$ takes one value for each of the $N$ experimental data $F_{exp,j}^{E}$, and $F_{cal,j}^{E}$ is the corresponding value of the excess property $F^{E}$ calculated from equation (5).

Values of $\varepsilon_{r}^{E}$, $n_{D}^{E}$ and $\left(\partial \varepsilon_{r}^{E}/\partial T\right)_{p}$ versus $\phi_{1}$ of 1-alkanol + amine systems at 298.15 K are plotted in Figures 1, 2 and 3 respectively with their corresponding Redlich-Kister regressions. Data on $n_{D}$ are plotted in Figure S1.

## 4. Discussion

Unless stated otherwise, the below values of the thermophysical properties and their corresponding excess functions are referred to $T = 298.15$ K and $\phi_{1} = 0.5$. On the other hand, $n$ stands for the number of C atoms of the 1-alkanol.

### 4.1. Excess relative permittivities

The rupture of interactions between molecules of the same species along mixing is associated to a negative contribution to $\varepsilon_r^E$. The creation of new interactions in the mixture forms multimers, whose total dipole moment can be more or less effective –in its impact on the macroscopic dipole moment under the action of an electric field– than in the ideal mixture. In the first case, the contribution to $\varepsilon_r^E$ is positive, whereas in the second case it is negative. 1-Alkanol + heptane mixtures show rather large negative values of this quantity (Figure 4): $\varepsilon_r^E$ = –1.075 ($n$ = 3), –2.225 ($n$ = 4), –2.525 ($n$ = 5), –2.875 ($n$ = 7), –1.775 ($n$ = 10) [12, 38-40]. These negative values can be attributed to the breaking of 1-alkanol self-association. For methanol, there exists a partial immiscibility region [41]. The $\varepsilon_r(n)$ variation follows the sequence: 1-propanol > 1-butanol > 1-pentanol > 1-heptanol < 1-decanol. It can be explained in terms of the lower and weaker self-association of longer 1-alkanols [19]. This statement also applies for the relative variation of $\varepsilon_r^E$ in the mixtures under study (Figures 1 and 4): 1.480 ($n$ = 1), –0.960 ($n$ = 3), –1.424 ($n$ = 4), –1.530 ($n$ = 5), –1.295 ($n$ = 7). These results are higher than those of heptane mixtures. This suggests that the formation of (1-alkanol)-HxA interactions yields a positive contribution to $\varepsilon_r^E$.

#### 4.1.1. Effect of cyclization

Cyclohexylamine (c-HxA) is a cyclic primary amine with a slightly higher permittivity than HxA ($\varepsilon_r^* = 4.53$ [19]). The trend observed in the two series of systems 1-alkanol + heptane and + HxA is the same as for 1-alkanol + c-HxA mixtures (Figure 4), for the same reasons: $\varepsilon_r^E$ = 2.218 ($n$ = 1 [13]), –0.269 ($n$ = 3 [19]), –0.848 ($n$ = 4 [19]), –0.915 ($n$ = 7 [19]) –0.411 ($n$ = 10 [19]). Therefore, cyclization of the amine leads to increased $\varepsilon_r^E$ values compared to those of systems with HxA; i.e., multimers formed by unlike molecules contribute more positively to $\varepsilon_r^E$ in cyclohexylamine solutions.

#### 4.1.2. Effect of aromaticity

The effect of aromaticity is more dramatic than that of cyclization. In fact, aniline ($\varepsilon_r^* = 7.004$ [42]) shows a greater value of the relative permittivity, underlining the importance of aniline-aniline interactions and the polarizability of the aromatic ring. The values of the corresponding excess property are of course negative [42] (Figure 4): –0.775 ($n$ = 1), –1.854 ($n$ = 3), –2.084 ($n$ = 5). In addition, they are lower than those of the mixtures with HxA or c-HxA. This may be explained taking into account that the breaking of the dipolar interactions between aniline molecules contributes more negatively to $\varepsilon_r^E$.

### 4.2. Entropy change with the electric field

$\varepsilon_r$ is a collective property and its magnitude in a liquid depends on its structure, the permanent dipole moment of its molecules and their polarizability. It must be highlighted that it is also affected by volume effects. In fact, let $\vec{B}$ denote the macroscopic dipole moment and $V$ the volume. The polarization (macroscopic dipole moment per unit volume) of the liquid, $\vec{B}/V$, is related to the intensity of the electric field, $\vec{E}$, through the equation $\vec{B}/V = (\varepsilon_r - 1)\varepsilon_0 \vec{E}$ ($\varepsilon_0$ = vacuum permittivity). In order to compare the response of different liquids to an electric field, it is desirable to work with the molar susceptibility, $\chi_m = (\varepsilon_r - 1)V_m$. This quantity, for a given electric field, is proportional to the macroscopic dipole moment resulting from a fixed amount (1 mol) of molecules. For a linear, isotropic and homogeneous dielectric at constant composition, the molar macroscopic dipole moment $B_m$ is related to $\chi_m$ and $E$ by:

$$\frac{1}{\varepsilon_0 E} B_m = \chi_m \qquad (7)$$

The $T$-dependence of $\chi_m$ is linked to the change of the molar entropy, $S_m$, with a variation of $E$ through the Schwarz relation ($p$ = pressure):

$$\frac{1}{\varepsilon_0 E}\left(\frac{\partial S_m}{\partial E}\right)_{T,p} = \left(\frac{\partial S_m}{\partial(\varepsilon_0 E^2/2)}\right)_{T,p} = \left(\frac{\partial \chi_m}{\partial T}\right)_p \qquad (8)$$

This variation is usually negative in common liquids like the ones considered in this work (Table 6), as it is associated with structure creation (dipolar ordering) by an increase of the electric field and a consequent negative variation of the entropy. It can be calculated from linear regressions of $\chi_m$ values in the temperature range (293.15 – 303.15) K. In the following discussion, we will use the notation $\eta_{m,p} = -\left(\frac{\partial \chi_m}{\partial T}\right)_p$.

The values of $\eta_{m,p}^*$ for the pure 1-alkanols increase with $n$ (Table 6). It indicates that a high self-association (in the absence of an electric field) decreases the ability of the electric field to create structure by orientating the dipoles of individual molecules, as the multimers present in the liquid are more stable and the rotational degrees of freedom are more constricted. The lower $\eta_{m,p}^*$ values of the amines seems to be due to their lower $\mu$. As HxA and c-HxA have similar dipole moments (Table 6), the fact that $\eta_{m,p}^*$(HxA) < $\eta_{m,p}^*$(c-HxA) indicates more freedom of rotation of the c-HxA molecules. Also, $\eta_{m,p}^*$(c-HxA) < $\eta_{m,p}^*$(aniline), which can be ascribed to an extra contribution to the polarizability of the molecules of aniline due to the presence of the aromatic ring.

It is interesting to analyse the $\eta_{m,p}(\phi_1)$ curves for 1-alkanol + HxA systems (Figure 5). The necessary volumetric properties to compute them have been taken from a previous work [4]. Due to the absence of experimental volumetric data in the whole range of temperature necessary for the complete set of studied systems, we have neglected the contribution from the temperature dependence of the excess molar volume; this approximation does not appreciably affect the $\eta_{m,p}$ results, as can be seen by performing the exact calculation for the system 1-butanol + HxA [43]. At low HxA concentrations, $\eta_{m,p}$ varies more rapidly when the 1-alkanol is longer. This shows that, in this region, the difficulty for the electric field to rotate the dipoles decreases more rapidly with concentration when the 1-alkanol self-association is lower and weaker.

### 4.3. Molar refraction

The refractive index at optical wavelengths is closely related to dispersion forces, since the molar refraction (or molar refractivity), $R_m$, defined by the Lorentz-Lorenz equation [29, 44]:

$$R_m = \frac{n_D^2 - 1}{n_D^2 + 2} V_m = \frac{N_A \alpha_e}{3\varepsilon_0} \quad (9)$$

(where $N_A$ and $\varepsilon_0$ stand for Avogadro's constant and the vacuum permittivity, respectively) is proportional to the mean electronic contribution, $\alpha_e$, to the polarizability, [29]. For the investigated systems, the values of $R_m / \text{cm}^3 \cdot \text{mol}^{-1}$ at $x_1 = 0.5$ are (Figure S2): 20.5 ($n = 1$), 25.2 ($n = 3$), 27.5 ($n = 4$), 29.8 ($n = 5$), 34.5 ($n = 7$). It is clear that dispersive interactions are more important in longer 1-alkanols. We have calculated the corresponding excess values, $R_m^E = R_m - R_m^{id}$, with $R_m^{id}$ evaluated substituting ideal values in equation (9). The curves are negative, which means a loss in dispersive interactions along mixing with respect to the ideal state, in which dipoles of different components do not interact. The minimum values occur at $x_1 \approx 0.5$; in the same units: –0.37 ($n = 1$), –0.28 ($n \geq 3$). The lower value of the methanol system can be ascribed to a larger number of hydrogen bonds formed by the two species along the mixing process.

### 4.4. Kirkwood-Fröhlich model

Some relevant hypotheses of the model are: (i) a molecule of a given polar compound is modelled as a dipole moment inside a spherical cavity; (ii) the effect of the induced polarization of the molecules is treated in macroscopic way, assuming that the dipole is rigid (it only rotates) and the cavity is filled by a continuous medium of relative permittivity $\varepsilon_r^\infty$ (the value of the permittivity at a high frequency at which only the induced polarizability contributes); (iii) long-range interactions are considered macroscopically by assuming that the outside of the cavity is a continuous dielectric of permittivity $\varepsilon_r$ ; (iv) short-range interactions are not neglected, and they are brought on stage by means of the so-called Kirkwood correlation factor, $g_K$, which provides information of the deviations from randomness of the orientation of a dipole with respect to its neighbours. This is an important parameter, as it provides information about specific interactions in the liquid state. For a mixture, $g_K$ can be determined, in the context of a one-fluid model [26], from macroscopic physical properties according to the expression [26-29]:

$$g_K = \frac{9k_B T V_m \varepsilon_0 (\varepsilon_r - \varepsilon_r^\infty)(2\varepsilon_r + \varepsilon_r^\infty)}{N_A \mu^2 \varepsilon_r (\varepsilon_r^\infty + 2)^2} \quad (10)$$

Here, $k_B$ is Boltzmann's constant; $N_A$, Avogadro's constant; $\varepsilon_0$, the vacuum permittivity; and $V_m$, the molar volume of the liquid at the working temperature, $T$. For polar compounds, $\varepsilon_r^\infty$ is estimated from the relation $\varepsilon_r^\infty = 1.1 n_D^2$ [45]. $\mu$ represents the dipole moment of the solution, estimated from the equation [26]:

$$\mu^2 = x_1 \mu_1^2 + x_2 \mu_2^2 \quad (11)$$

where $\mu_i$ stands for the dipole moment of component i (=1,2). Calculations have been conducted using smoothed values of $V_m^E$ [4], $n_D^E$ (this work) and $\varepsilon_r^E$ (this work) at $\Delta x_1 = 0.01$. The source and values of $\mu_i$ are collected in Table 6.

Our calculations on $g_K$ curves for 1-alkanol + HxA systems can be seen in Figure 6. They support the conclusions extracted from the analysis of $\eta_{m,p}$, indicating that the mixture structure varies very rapidly with the HxA concentration for $n$ = 3, 4, 5, 7. In contrast, for the methanol system $g_K$ changes slowly from $\phi_1 > 0.6$ approximately, and this indicates that HxA is not able to break effectively the methanol self-association at such concentrations. This phenomenon remarks the strong relationship between the magnitude of the rupture of the 1-alkanol self-association by the amine and the dielectric behaviour of the mixtures.

We have also evaluated the excess Kirkwood correlation factors, $g_K^E = g_K - g_K^{id}$, where $g_K^{id}$ is calculated substituting the real quantities by ideal ones in equation (10). The values for 1-alkanol + HxA systems are (Figures 7 and 8): 0.170 ($n$ = 1), –0.257 ($n$ = 3), –0.421 ($n$ = 4), –0.505 ($n$ = 5), –0.508 ($n$ = 7). The positive value for the methanol mixture can be justified by the formation of strong methanol-HxA interactions, which is consistent with the above analyses. The minima of the $g_K^E$ curves occurs at lower $\phi_1$ than in the $\varepsilon_r^E$ curves. For the minimum of the curves, $g_K^E$(1-pentanol) > $g_K^E$(1-heptanol), while the opposite behaviour is encountered for $\varepsilon_r^E$. Thus, according to the Kirkwood-Fröhlich model, the destruction of the correlations of the dipoles is not the only responsible for the $\varepsilon_r^E$ minima, but there are other effects involved. For c-HxA systems, $g_K^E$ values are higher (Figure 8), indicating that in these mixtures the balance of destruction and creation of correlations is more inclined to the latter than in the case 1-alkanol + HxA. Aniline systems are quite interesting, as $g_K^E$(HxA) < $g_K^E$(aniline) for the 1-pentanol mixtures (Figure 8). This phenomenon may be related to the higher importance of the rupture of interactions between like molecules in 1-alkanol + aniline solutions, as showed by $\varepsilon_r^E$ values and also $H_m^E$ (see introduction).

## 5. Conclusions

$\varepsilon_r$ and $n_D$ measurements have been reported for the 1-alkanol + n-hexylamine systems at (293.15-303.15) K. The formation of multimers built by unlike molecules contributes positively to $\varepsilon_r^E$. Such contribution is dominant for the methanol mixture and $\varepsilon_r^E$ is positive. For the remaining systems, the dominant contributions arise from the breaking of interactions between like molecules, and $\varepsilon_r^E$ values are negative. For a given 1-alkanol, $\varepsilon_r^E$ changes in the sequence: cyclohexylamine > n-hexylamine > aniline. The application of the Kirkwood-Fröhlich model confirms these findings. Calculations on $R_m$ show that dispersive interactions in the studied mixtures increase with the length of the 1-alkanol.

## Acknowledgements

F. Hevia and A. Cobos are grateful to Ministerio de Educación, Cultura y Deporte for the grants FPU14/04104 and FPU15/05456 respectively. The authors gratefully acknowledge the financial support received from the Consejería de Educación y Cultura of Junta de Castilla y León, under Project BU034U16.

## References

[1] A. Heintz, P.K. Naicker, S.P. Verevkin, R. Pfestorf, Thermodynamics of alkanol + amine mixtures. Experimental results and ERAS model calculations of the heat of mixing, Ber. Bunsenges. Phys. Chem., 102 (1998) 953-959. DOI: 10.1002/bbpc.19981020707.

[2]   K. Nakanishi, H. Touhara, N. Watanabe, Studies on Associated Solutions. II. Heat of Mixing of Methanol with Aliphatic Amines, Bull. Chem. Soc. Jpn., 43 (1970) 2671-2676. DOI: 10.1246/bcsj.43.2671.

[3]   A. Heintz, A New Theoretical Approach for Predicting Excess Properties of Alkanol/Alkane Mixtures, Ber. Bunsenges. Phys. Chem., 89 (1985) 172-181. DOI: 10.1002/bbpc.19850890217.

[4]   S. Villa, N. Riesco, I. García de la Fuente, J.A. González, J.C. Cobos, Thermodynamics of mixtures with strongly negative deviations from Raoult's law. Part 8. Excess molar volumes at 298.15 K for 1-alkanol + isomeric amine (C6H15N) systems: Characterization in terms of the ERAS model, Fluid Phase Equilib., 216 (2004) 123-133. DOI: 10.1016/j.fluid.2003.10.008.

[5]   J.A. González, I. García de la Fuente, J.C. Cobos, Thermodynamics of mixtures with strongly negative deviations from Raoult's Law: Part 4. Application of the DISQUAC model to mixtures of 1-alkanols with primary or secondary linear amines. Comparison with Dortmund UNIFAC and ERAS results, Fluid Phase Equilib., 168 (2000) 31-58. DOI: 10.1016/S0378-3812(99)00326-X.

[6]   R. Srivastava, B.D. Smith, Total-pressure vapor-liquid equilibrium data for binary systems of diethylamine with acetone, acetonitrile, and methanol, J. Chem. Eng. Data, 30 (1985) 308-313. DOI: 10.1021/je00041a022.

[7]   L. Wang, G.C. Benson, B.C.Y. Lu, Excess enthalpies of 1-propanol + n-hexane + n-decane or n-dodecane at 298.15 K, J. Chem. Eng. Data, 37 (1992) 403-406. DOI: 10.1021/je00008a007.

[8]   S.-c. Hwang, R.L. Robinson, Vapor-liquid equilibria at 25 ºC for nine alcohol-hydrocarbon binary systems, J. Chem. Eng. Data, 22 (1977) 319-325. DOI: 10.1021/je60074a025.

[9]   S. Villa, N. Riesco, I. García de la Fuente, J.A. González, J.C. Cobos, Thermodynamics of mixtures with strongly negative deviations from Raoult's law: Part 5. Excess molar volumes at 298.15 K for 1-alkanols+dipropylamine systems: characterization in terms of the ERAS model, Fluid Phase Equilib., 190 (2001) 113-125. DOI: 10.1016/S0378-3812(01)00595-7.

[10]  S. Villa, N. Riesco, I. García de la fuente, J.A. González, J.C. Cobos, Thermodynamics of mixtures with strongly negative deviations from Raoult's law: Part 6. Excess molar volumes at 298.15 K for 1-alkanols + dibutylamine systems. Characterization in terms

of the ERAS model, Fluid Phase Equilib., 198 (2002) 313-329. DOI: 10.1016/S0378-3812(01)00808-1.

[11] L.F. Sanz, J.A. González, I. García De La Fuente, J.C. Cobos, Thermodynamics of mixtures with strongly negative deviations from Raoult's law. XI. Densities, viscosities and refractives indices at (293.15–303.15) K for cyclohexylamine + 1-propanol, or +1-butanol systems, J. Mol. Liq., 172 (2012) 26-33. DOI: 10.1016/j.molliq.2012.05.003.

[12] L.F. Sanz, J.A. González, I. García de la Fuente, J.C. Cobos, Thermodynamics of mixtures with strongly negative deviations from Raoult's law. XII. Densities, viscosities and refractive indices at T = (293.15 to 303.15) K for (1-heptanol, or 1-decanol + cyclohexylamine) systems. Application of the ERAS model to (1-alkanol + cyclohexylamine) mixtures, J. Chem. Thermodyn., 80 (2015) 161-171. DOI: 10.1016/j.jct.2014.09.005.

[13] L.F. Sanz, J.A. González, I.G. De La Fuente, J.C. Cobos, Thermodynamics of mixtures with strong negative deviations from raoult's law. XIV. density, permittivity, refractive index and viscosity data for the methanol + cyclohexylamine mixture at (293.15–303.15) K, Thermochim. Acta, 631 (2016) 18-27. DOI: 10.1016/j.tca.2016.03.002.

[14] U. Domańska, M. Głoskowska, Experimental Solid + Liquid Equilibria and Excess Molar Volume of Alkanol + Octylamine Mixtures. Analysis in Terms of ERAS, DISQUAC, and Modified UNIFAC, J. Chem. Eng. Data, 49 (2004) 101-108. DOI: 10.1021/je0301895.

[15] K. Nakanishi, H. Touhara, Excess molar enthalpies of (methanol + aniline), (methanol + N-methylaniline), and (methanol + N,N-dimethylaniline), J. Chem. Thermodyn., 18 (1986) 657-660. DOI: 10.1016/0021-9614(86)90067-4.

[16] I. Nagata, Excess enthalpies of (aniline + butan-1-ol) and of (aniline + butan-1-ol + benzene) at the temperature 298.15 K, J. Chem. Thermodyn., 25 (1993) 1281-1285. DOI: 10.1006/jcht.1993.1127.

[17] H. Matsuda, K. Ochi, K. Kojima, Determination and Correlation of LLE and SLE Data for the Methanol + Cyclohexane, Aniline + Heptane, and Phenol + Hexane System, J. Chem. Eng. Data, 48 (2003) 184-189. DOI: 10.1021/je020156+.

[18] S. Villa, R. Garriga, P. Pérez, M. Gracia, J.A. González, I.G. de la Fuente, J.C. Cobos, Thermodynamics of mixtures with strongly negative deviations from Raoult's law: Part 9. Vapor–liquid equilibria for the system 1-propanol + di-n-propylamine at six temperatures between 293.15 and 318.15 K, Fluid Phase Equilib., 231 (2005) 211-220. DOI: 10.1016/j.fluid.2005.01.013.

[19] J.A. González, L.F. Sanz, I. García de la Fuente, J.C. Cobos, Thermodynamics of mixtures with strong negative deviations from Raoult's law. XIII. Relative permittivities for (1-alkanol + cyclohexylamine) systems, and dielectric study of (1-alkanol + polar) compound (amine, amide or ether) mixtures, J. Chem. Thermodyn., 91 (2015) 267-278. DOI: 10.1016/j.jct.2015.07.032.

[20] J.A. González, I.G. de la Fuente, J.C. Cobos, Thermodynamics of mixtures with strongly negative deviations from Raoult's law. Part 3. Application of the DISQUAC model to mixtures of triethylamine with alkanols. Comparison with Dortmund UNIFAC and ERAS results, Can. J. Chem., 78 (2000) 1272-1284. DOI: 10.1139/v00-114.

[21] J.A. González, I. Mozo, I. García de la Fuente, J.C. Cobos, Thermodynamics of organic mixtures containing amines. IV. Systems with aniline, Can. J. Chem., 83 (2005) 1812-1825. DOI: 10.1139/v05-190.

[22] J.A. González, I. Mozo, I.G.d.l. Fuente, J.C. Cobos, Thermodynamics of organic mixtures containing amines: V. Systems with pyridines, Thermochim. Acta, 441 (2006) 53-68. DOI: 10.1016/j.tca.2005.11.027.

[23] J.A. González, I. Mozo, I.G. de la Fuente, J.C. Cobos, N. Riesco, Thermodynamics of mixtures containing amines: VII. Systems containing dimethyl or trimethylpyridines, Thermochim. Acta, 467 (2008) 30-43. DOI: 10.1016/j.tca.2007.10.011.

[24] J.A. González, I.G. de la Fuente, I. Mozo, J.C. Cobos, N. Riesco, Thermodynamics of Organic Mixtures Containing Amines. VII. Study of Systems Containing Pyridines in Terms of the Kirkwood−Buff Formalism, Ind. Eng. Chem. Res., 47 (2008) 1729-1737. DOI: 10.1021/ie071226e.

[25] J.A. González, J.C. Cobos, I. García de la Fuente, I. Mozo, Thermodynamics of mixtures containing amines. IX. Application of the concentration–concentration structure factor to the study of binary mixtures containing pyridines, Thermochim. Acta, 494 (2009) 54-64. DOI: 10.1016/j.tca.2009.04.017.

[26] J.C.R. Reis, T.P. Iglesias, Kirkwood correlation factors in liquid mixtures from an extended Onsager-Kirkwood-Frohlich equation, Phys. Chem. Chem. Phys., 13 (2011) 10670-10680. DOI: 10.1039/C1CP20142E.

[27] H. Fröhlich, Theory of Dielectrics, Clarendon Press, Oxford, 1958.

[28] C. Moreau, G. Douhéret, Thermodynamic and physical behaviour of water + acetonitrile mixtures. Dielectric properties, J. Chem. Thermodyn., 8 (1976) 403-410. DOI: 10.1016/0021-9614(76)90060-4.

[29] A. Chelkowski, Dielectric Physics, Elsevier, Amsterdam, 1980.

[30]  CIAAW, Atomic weights of the elements 2015, ciaaw.org/atomic-weights.htm (accessed 2015).

[31]  J.A. González, I. Alonso, I. Mozo, I. García de la Fuente, J.C. Cobos, Thermodynamics of (ketone + amine) mixtures. Part VI. Volumetric and speed of sound data at (293.15, 298.15, and 303.15) K for (2-heptanone + dipropylamine, +dibutylamine, or +triethylamine) systems, J. Chem. Thermodyn., 43 (2011) 1506-1514. DOI: 10.1016/j.jct.2011.05.003.

[32]  K.N. Marsh, Recommended reference materials for the realization of physicochemical properties, Blackwell Scientific Publications, Oxford, UK, 1987.

[33]  V. Alonso, J.A. González, I. García de la Fuente, J.C. Cobos, Dielectric and refractive index measurements for the systems 1-pentanol + octane, or + dibutyl ether or for dibutyl ether + octane at different temperatures, Thermochim. Acta, 543 (2012) 246-253. DOI: 10.1016/j.tca.2012.05.036.

[34]  J.C.R. Reis, T.P. Iglesias, G. Douhéret, M.I. Davis, The permittivity of thermodynamically ideal liquid mixtures and the excess relative permittivity of binary dielectrics, Phys. Chem. Chem. Phys., 11 (2009) 3977-3986. DOI: 10.1039/B820613A.

[35]  J.C.R. Reis, I.M.S. Lampreia, Â.F.S. Santos, M.L.C.J. Moita, G. Douhéret, Refractive Index of Liquid Mixtures: Theory and Experiment, ChemPhysChem, 11 (2010) 3722-3733. DOI: 10.1002/cphc.201000566.

[36]  O. Redlich, A.T. Kister, Algebraic Representation of Thermodynamic Properties and the Classification of Solutions, Ind. & Eng. Chem., 40 (1948) 345-348. DOI: 10.1021/ie50458a036.

[37]  P.R. Bevington, D.K. Robinson, Data Reduction and Error Analysis for the Physical Sciences, McGraw-Hill, New York, 2000.

[38]  N.V. Sastry, M.K. Valand, Densities, Speeds of Sound, Viscosities, and Relative Permittivities for 1-Propanol + and 1-Butanol + Heptane at 298.15 K and 308.15 K, J. Chem. Eng. Data, 41 (1996) 1421-1425. DOI: 10.1021/je960135d.

[39]  N.V. Sastry, M.K. Valand, Dielectric constants, refractive indexes and polarizations for 1-Alcohol +Heptane mixtures at 298.15 and 308.15 K, Ber. Bunsenges. Phys. Chem., 101 (1997) 243-250. DOI: 10.1002/bbpc.19971010212.

[40]  J.A. Riddick, W.B. Bunger, T.K. Sakano, Organic solvents: physical properties and methods of purification, Wiley, New York, 1986.

[41]   A. Skrzecz, Critical evaluation of solubility data in binary systems formed by methanol with n-hydrocarbons, Thermochim. Acta, 182 (1991) 123-131. DOI: 10.1016/0040-6031(91)87013-M.

[42]   V. Alonso, Estudio experimental de propiedades termofísicas de mezclas binarias formadas por 1-alcohol + alcano, + éter lineal o + amina aromática primaria, in: Departamento de Física Aplicada, Facultad de Ciencias, Universidad de Valladolid, 2016.

[43]   I.R. Radović, M.L. Kijevčanin, S.P. Šerbanović, B.D. Djordjević, 1-Butanol + hexylamine + $n$-heptane at temperature range (288.15–323.15 K): Experimental density data, excess molar volumes determination and modeling with cubic EOS, Fluid Phase Equilib., 298 (2010) 117-130. DOI: 10.1016/j.fluid.2010.07.011.

[44]   P. Brocos, A. Piñeiro, R. Bravo, A. Amigo, Refractive indices, molar volumes and molar refractions of binary liquid mixtures: concepts and correlations, Phys. Chem. Chem. Phys., 5 (2003) 550-557. DOI: 10.1039/B208765K.

[45]   Y. Marcus, The structuredness of solvents, J. Solution Chem., 21 (1992) 1217-1230. DOI: 10.1007/bf00667218.

[46]   R.D. Bezman, E.F. Casassa, R.L. Kay, The temperature dependence of the dielectric constants of alkanols, 73–74 (1997) 397-402. DOI: 10.1016/S0167-7322(97)00082-2.

[47]   J. Canosa, A. Rodríguez, J. Tojo, Binary mixture properties of diethyl ether with alcohols and alkanes from 288.15 K to 298.15 K, Fluid Phase Equilib., 156 (1999) 57-71. DOI: 10.1016/S0378-3812(99)00032-1.

[48]   G.J. Janz, R.P.T. Tomkins, Nonaqueus Electrolytes Handbook, Vol. 1, Academic Press, New York, 1972.

[49]   S.P. Serbanovic, M.L. Kijevcanin, I.R. Radovic, B.D. Djordjevic, Effect of temperature on the excess molar volumes of some alcohol + aromatic mixtures and modelling by cubic EOS mixing rules, Fluid Phase Equilib., 239 (2006) 69-82. DOI: 10.1016/j.fluid.2005.10.022.

[50]   S. Chen, Q. Lei, W. Fang, Density and Refractive Index at 298.15 K and Vapor−Liquid Equilibria at 101.3 kPa for Four Binary Systems of Methanol, n-Propanol, n-Butanol, or Isobutanol with N-Methylpiperazine, J. Chem. Eng. Data, 47 (2002) 811-815. DOI: 10.1021/je010249b.

[51]   M.I. Aralaguppi, C.V. Jadar, T.M. Aminabhavi, Density, Viscosity, Refractive Index, and Speed of Sound in Binary Mixtures of Acrylonitrile with Methanol, Ethanol,

Propan-1-ol, Butan-1-ol, Pentan-1-ol, Hexan-1-ol, Heptan-1-ol, and Butan-2-ol, J. Chem. Eng. Data, 44 (1999) 216-221. DOI: 10.1021/je9802219.

[52] R. Anwar Naushad, S. Yasmeen, Volumetric, compressibility and viscosity studies of binary mixtures of [EMIM][NTf2] with ethylacetate/methanol at (298.15–323.15) K, J. Mol. Liq., 224, Part A (2016) 189-200. DOI: 10.1016/j.molliq.2016.09.077.

[53] S.M. Pereira, T.P. Iglesias, J.L. Legido, L. Rodríguez, J. Vijande, Changes of refractive index on mixing for the binary mixtures {xCH3OH+(1−x)CH3OCH2(CH2OCH2)3CH2OCH3} and {xCH3OH+(1−x)CH3OCH2(CH2OCH2)nCH2OCH3} (n=3–9) at temperatures from 293.15 K to 333.15 K, J. Chem. Thermodyn., 30 (1998) 1279-1287. DOI: 10.1006/jcht.1998.0395.

[54] A. Rodríguez, J. Canosa, J. Tojo, Density, Refractive Index, and Speed of Sound of Binary Mixtures (Diethyl Carbonate + Alcohols) at Several Temperatures, J. Chem. Eng. Data, 46 (2001) 1506-1515. DOI: 10.1021/je010148d.

[55] A.P. Gregory, R.N. Clarke, Traceable measurements of the static permittivity of dielectric reference liquids over the temperature range 5–50 °C, Meas. Sci. Technol., 16 (2005) 1506. DOI: 10.1088/0957-0233/16/7/013.

[56] M.J. Fontao, M. Iglesias, Effect of Temperature on the Refractive Index of Aliphatic Hydroxilic Mixtures (C2–C3), Int. J. Thermophys., 23 (2002) 513-527. DOI: 10.1023/A:1015113604024.

[57] J.L. Hales, J.H. Ellender, Liquid densities from 293 to 490 K of nine aliphatic alcohols, J. Chem. Thermodyn., 8 (1976) 1177-1184. DOI: 10.1016/0021-9614(76)90126-9.

[58] N.G. Tsierkezos, I.E. Molinou, A.C. Filippou, Thermodynamic Properties of Binary Mixtures of Cyclohexanone with n-Alkanols (C1–C5) at 293.15 K, J. Solution Chem., 34 (2005) 1371-1386. DOI: 10.1007/s10953-005-8508-9.

[59] C. Yang, H. Lai, Z. Liu, P. Ma, Density and Viscosity of Binary Mixtures of Diethyl Carbonate with Alcohols at (293.15 to 363.15) K and Predictive Results by UNIFAC-VISCO Group Contribution Method, J. Chem. Eng. Data, 51 (2006) 1345-1351. DOI: 10.1021/je0600808.

[60] C.P. Smyth, W.N. Stoops, The dielectric polarization of liquids. VI. Ethyl iodide, ethanol, normal-butanol and normal-octanol, J. Am. Chem. Soc., 51 (1929) 3312-3329. DOI: 10.1021/ja01386a019.

[61] B. Giner, A. Villares, M.C. López, F.M. Royo, C. Lafuente, Refractive indices and molar refractions for isomeric chlorobutanes with isomeric butanols, Phys. Chem. Liq., 43 (2005) 13-23. DOI: 10.1080/00319104200030518.

[62] E. Jiménez, M. Cabanas, L. Segade, S. García-Garabal, H. Casas, Excess volume, changes of refractive index and surface tension of binary 1,2-ethanediol + 1-propanol or 1-butanol mixtures at several temperatures, Fluid Phase Equilib., 180 (2001) 151-164. DOI: 10.1016/S0378-3812(00)00519-7.

[63] G.A. Iglesias-Silva, A. Guzmán-López, G. Pérez-Durán, M. Ramos-Estrada, Densities and Viscosities for Binary Liquid Mixtures of n-Undecane + 1-Propanol, + 1-Butanol, + 1-Pentanol, and + 1-Hexanol from 283.15 to 363.15 K at 0.1 MPa, J. Chem. Eng. Data, 61 (2016) 2682-2699. DOI: 10.1021/acs.jced.6b00121.

[64] T.P. Iglesias, J.L. Legido, S.M. Pereira, B. de Cominges, M.I. Paz Andrade, Relative permittivities and refractive indices on mixing for (n-hexane + 1-pentanol, or 1-hexanol, or 1-heptanol ) atT = 298.15 K, J. Chem. Thermodyn., 32 (2000) 923-930. DOI: 10.1006/jcht.2000.0661.

[65] M.N.M. Al-Hayan, Densities, excess molar volumes, and refractive indices of 1,1,2,2-tetrachloroethane and 1-alkanols binary mixtures, J. Chem. Thermodyn., 38 (2006) 427-433. DOI: 10.1016/j.jct.2005.06.015.

[66] S.P. Patil, A.S. Chaudhari, M.P. Lokhande, M.K. Lande, A.G. Shankarwar, S.N. Helambe, B.R. Arbad, S.C. Mehrotra, Dielectric Measurements of Aniline and Alcohol Mixtures at 283, 293, 303, and 313 K Using the Time Domain Technique, J. Chem. Eng. Data, 44 (1999) 875-878. DOI: 10.1021/je980250j.

[67] Á. Piñeiro, P. Brocos, A. Amigo, M. Pintos, R. Bravo, Refractive Indexes of Binary Mixtures of Tetrahydrofuran with 1-Alkanols at 25°C and Temperature Dependence of n and ρ for the Pure Liquids, J. Solution Chem., 31 (2002) 369-380. DOI: 10.1023/A:1015807331250.

[68] J.J. Cano-Gómez, G.A. Iglesias-Silva, E.O. Castrejón-González, M. Ramos-Estrada, K.R. Hall, Density and Viscosity of Binary Liquid Mixtures of Ethanol + 1-Hexanol and Ethanol + 1-Heptanol from (293.15 to 328.15) K at 0.1 MPa, J. Chem. Eng. Data, 60 (2015) 1945-1955. DOI: 10.1021/je501133u.

[69] U. Domańska, M. Królikowska, Density and Viscosity of Binary Mixtures of {1-Butyl-3-methylimidazolium Thiocyanate + 1-Heptanol, 1-Octanol, 1-Nonanol, or 1-Decanol}, J. Chem. Eng. Data, 55 (2010) 2994-3004. DOI: 10.1021/je901043q.

[70] C. Wohlfahrt, Static Dielectric Constants of Pure Liquids and Binary Liquid Mixtures. Landolt-Börnstein - Group IV Physical Chemistry Vol. 6, Springer Berlin Heidelberg, Berlin, 1991.

[71] S. Otín, J. Fernández, J.M. Embid, I. Velasco, C.G. Losa, Thermodynamic and Dielectric Properties of Binary Polar + Non-Polar Mixtures I. Static Dielectric Constants and Excess Molar Enthalpies of n-Alkylamine + n-Dodecane Systems, Ber. Bunsenges. Phys. Chem., 90 (1986) 1179-1183. DOI: 10.1002/bbpc.19860901212.

[72] C. Wohlfahrt, Optical Constants. Refractive Indices of Pure Liquids and Binary Liquid Mixtures. Landolt-Börnstein - Group III Condensed Matter Vol. 47, Springer Berlin Heidelberg, Berlin, 2008.

[73] Y. Miyake, A. Baylaucq, F. Plantier, D. Bessières, H. Ushiki, C. Boned, High-pressure (up to 140 MPa) density and derivative properties of some (pentyl-, hexyl-, and heptyl-) amines between (293.15 and 353.15) K, J. Chem. Thermodyn., 40 (2008) 836-845. DOI: 10.1016/j.jct.2008.01.006.

[74] P. Góralski, M. Wasiak, A. Bald, Heat Capacities, Speeds of Sound, and Isothermal Compressibilities of Some n-Amines and Tri-n-amines at 298.15 K, J. Chem. Eng. Data, 47 (2002) 83-86. DOI: 10.1021/je010206v.

[75] M. El-Hefnawy, K. Sameshima, T. Matsushita, R. Tanaka, Apparent Dipole Moments of 1-Alkanols in Cyclohexane and n-Heptane, and Excess Molar Volumes of (1-Alkanol + Cyclohexane or n-Heptane) at 298.15 K, J. Solution Chem., 34 (2005) 43-69. DOI: 10.1007/s10953-005-2072-1.

[76] A.L. McClellan, Tables of Experimental Dipole Moments, Vols. 1,2,3, Rahara Enterprises, El Cerrito, US, 1974.

[77] D.R. Lide, CRC Handbook of Chemistry and Physics, 90th Edition, CRC Press/Taylor and Francis, Boca Raton, FL, 2010.

[78] F. Hevia, J.A. González, A. Cobos, I. García de la Fuente, L.F. Sanz, Thermodynamics of amide + amine mixtures. 4. Relative permittivities of N,N-dimethylacetamide + N-propylpropan-1-amine, + N-butylbutan-1-amine, + butan-1-amine, or + hexan-1-amine systems and of N,N-dimethylformamide + aniline mixture at several temperatures. Characterization of amine + amide systems using ERAS, J. Chem. Thermodyn., 118 (2018) 175-187. DOI: 10.1016/j.jct.2017.11.011.

Table 1

Sample description.

| Chemical name | CAS Number | Source | Purification method | Purity[a] |
|---|---|---|---|---|
| methanol | 67-56-1 | Sigma-Aldrich | none | 99.99% |
| 1-propanol | 71-23-8 | Sigma-Aldrich | none | 99.84% |
| 1-butanol | 71-36-3 | Sigma-Aldrich | none | 99.86% |
| 1-pentanol | 71-41-0 | Sigma-Aldrich | none | 99.9% |
| 1-heptanol | 111-70-6 | Sigma-Aldrich | none | 99.8% |
| *n*-hexylamine (HxA) | 111-26-2 | Aldrich | none | 99.9% |

[a] In mole fraction. Provided by the supplier by gas chromatography.

Table 2

Relative permittivity at frequency $\nu = 1$ MHz, $\varepsilon_r^*$, refractive index, $n_D^*$, and density, $\rho^*$, of pure compounds at temperature $T$ and pressure $p = 0.1$ MPa. [a]

| Compound | T/K | $\varepsilon_r^*$ | | $n_D^*$ | | $\rho^*$ / g·cm$^{-3}$ | |
|---|---|---|---|---|---|---|---|
| | | Exp. | Lit. | Exp. | Lit. | Exp. | Lit. |
| methanol | 293.15 | 33.569 | 33.61 [46] | 1.32862 | 1.32859 [47] | 0.79163 | 0.7916 [48] |
| | | | | | | | 0.791400 [49] |
| | 298.15 | 32.619 | 32.62 [46] | 1.32654 | 1.32652 [50] | 0.78695 | 0.7869 [51] |
| | | | | | | | 0.786884 [52] |
| | 303.15 | 31.652 | 31.66 [46] | 1.32439 | 1.32457 [53] | 0.78222 | 0.782158 [52] |
| | | | | | 1.32410 [54] | | |
| 1-propanol | 293.15 | 21.150 | 21.15 [55] | 1.38514 | 1.38512 [56] | 0.80366 | 0.80361 [57] |
| | 298.15 | 20.449 | 20.42 [55] | 1.38306 | 1.38307 [54] | 0.79968 | 0.79960 [57] |
| | 303.15 | 19.784 | 19.75 [55] | 1.38102 | 1.38104 [54] | 0.79566 | 0.79561 [57] |
| 1-butanol | 293.15 | 18.192 | 18.19 [55] | 1.39931 | 1.3993 [58] | 0.80985 | 0.80982 [59] |
| | | | | | | | 0.8098 [60] |
| | 298.15 | 17.545 | 17.53 [55] | 1.39733 | 1.397336 [61] | 0.80606 | 0.80606 [59] |
| | 303.15 | 16.933 | 16.89 [55] | 1.39529 | 1.3953 [62] | 0.80222 | 0.8022 [60] |
| 1-pentanol | 293.15 | 15.701 | 15.63 [46] | 1.40985 | 1.40986 [54] | 0.81466 | 0.81468 [63] |
| | 298.15 | 15.102 | 15.08 [64] | 1.40793 | 1.40789 [54] | 0.81103 | 0.81103 [63] |
| | 303.15 | 14.536 | 14.44 [46] | 1.40590 | 1.40592 [65] | 0.80735 | 0.81737 [63] |
| 1-heptanol | 293.15 | 12.019 | 11.54 [66] | 1.42425 | 1.42433 [67] | 0.82237 | 0.8223 [68] |
| | 298.15 | 11.504 | 11.45 [64] | 1.42235 | 1.42240 [67] | 0.81890 | 0.81881 [69] |
| | 303.15 | 11.014 | 11.07 [70] | 1.42047 | 1.42047 [65] | 0.81537 | 0.8153 [68] |
| | | | | | 1.42048 [67] | | |
| HxA | 293.15 | 3.964 | 3.94 [71] | 1.41808 | 1.4180 [72] | 0.76443 | 0.7651 [73] |
| | 298.15 | 3.904 | | 1.41563 | 1.41550 [72] | 0.76019 | 0.76013 [74] |
| | 303.15 | 3.846 | 3.83 [71] | 1.41321 | 1.4131 [72] | 0.75590 | 0.7562 [73] |

[a]The standard uncertainties are: $u(T) = 0.02$ K (for $\rho^*$ measurements, $u(T) = 0.01$ K); $u(p) = 1$ kPa; $u(\nu) = 20$ Hz; $u(n_D^*) = 0.00008$. The relative standard uncertainties are: $u_r(\rho^*) = 0.0012$, $u_r(\varepsilon_r^*) = 0.003$.

Table 3

Volume fractions of 1-alkanol, $\phi_1$, relative permittivities, $\varepsilon_r$, and excess relative permittivities, $\varepsilon_r^E$, of 1-alkanol (1) + HxA (2) mixtures as functions of the mole fraction of the 1-alkanol, $x_1$, at temperature $T$, pressure $p = 0.1$ MPa and frequency $\nu = 1$ MHz. [a]

| $x_1$ | $\phi_1$ | $\varepsilon_r$ | $\varepsilon_r^E$ | $x_1$ | $\phi_1$ | $\varepsilon_r$ | $\varepsilon_r^E$ |
|---|---|---|---|---|---|---|---|
| methanol (1) + HxA (2) ; $T$/K = 293.15 ||||||||
| 0.0000 | 0.0000 | 3.964 |         | 0.7016 | 0.4182 | 17.646 | 1.301 |
| 0.0534 | 0.0170 | 4.367 | − 0.100 | 0.8002 | 0.5505 | 21.864 | 1.602 |
| 0.1220 | 0.0408 | 4.964 | − 0.208 | 0.8485 | 0.6313 | 24.251 | 1.597 |
| 0.1906 | 0.0672 | 5.681 | − 0.272 | 0.8984 | 0.7300 | 26.928 | 1.352 |
| 0.3081 | 0.1198 | 7.244 | − 0.267 | 0.9496 | 0.8521 | 30.037 | 0.847 |
| 0.3950 | 0.1664 | 8.783 | − 0.107 | 0.9834 | 0.9477 | 32.343 | 0.322 |
| 0.4982 | 0.2329 | 11.109 | 0.250 | 1.0000 | 1.0000 | 33.569 |       |
| 0.6035 | 0.3176 | 14.210 | 0.843 |        |        |        |       |
| methanol (1) + HxA (2) ; $T$/K = 298.15 ||||||||
| 0.0000 | 0.0000 | 3.904 |         | 0.7016 | 0.4183 | 17.127 | 1.212 |
| 0.0534 | 0.0170 | 4.293 | − 0.099 | 0.8002 | 0.5506 | 21.254 | 1.540 |
| 0.1220 | 0.0408 | 4.870 | − 0.206 | 0.8485 | 0.6314 | 23.597 | 1.562 |
| 0.1906 | 0.0672 | 5.564 | − 0.270 | 0.8984 | 0.7301 | 26.203 | 1.334 |
| 0.3081 | 0.1199 | 7.085 | − 0.262 | 0.9496 | 0.8521 | 29.216 | 0.844 |
| 0.3950 | 0.1665 | 8.557 | − 0.128 | 0.9834 | 0.9477 | 31.440 | 0.323 |
| 0.4982 | 0.2329 | 10.806 | 0.214 | 1.0000 | 1.0000 | 32.619 |       |
| 0.6035 | 0.3177 | 13.794 | 0.767 |        |        |        |       |
| methanol (1) + HxA (2) ; $T$/K = 303.15 ||||||||
| 0.0000 | 0.0000 | 3.964 |         | 0.7016 | 0.4182 | 17.646 | 1.301 |
| 0.0534 | 0.0170 | 4.367 | − 0.100 | 0.8002 | 0.5505 | 21.864 | 1.602 |
| 0.1220 | 0.0408 | 4.964 | − 0.208 | 0.8485 | 0.6313 | 24.251 | 1.597 |
| 0.1906 | 0.0672 | 5.681 | − 0.272 | 0.8984 | 0.7300 | 26.928 | 1.352 |
| 0.3081 | 0.1198 | 7.244 | − 0.267 | 0.9496 | 0.8521 | 30.037 | 0.847 |
| 0.3950 | 0.1664 | 8.783 | − 0.107 | 0.9834 | 0.9477 | 32.343 | 0.322 |
| 0.4982 | 0.2329 | 11.109 | 0.250 | 1.0000 | 1.0000 | 33.569 |       |
| 0.6035 | 0.3176 | 14.210 | 0.843 |        |        |        |       |
| 1-propanol (1) + HxA (2) ; $T$/K = 293.15 ||||||||
| 0.0000 | 0.0000 | 3.964 |         | 0.6097 | 0.4688 | 10.990 | − 1.031 |
| 0.0708 | 0.0413 | 4.411 | − 0.263 | 0.6977 | 0.5659 | 12.847 | − 0.843 |
| 0.1073 | 0.0636 | 4.663 | − 0.394 | 0.8044 | 0.6991 | 15.469 | − 0.510 |
| 0.1470 | 0.0887 | 4.953 | − 0.535 | 0.8406 | 0.7487 | 16.431 | − 0.400 |
| 0.1935 | 0.1194 | 5.330 | − 0.686 | 0.8989 | 0.8340 | 18.080 | − 0.217 |
| 0.3070 | 0.2002 | 6.436 | − 0.969 | 0.9504 | 0.9154 | 19.605 | − 0.091 |
| 0.3941 | 0.2687 | 7.471 | − 1.111 | 1.0000 | 1.0000 | 21.150 | |
| 0.5056 | 0.3662 | 9.113 | − 1.145 | | | | |

1-propanol (1) + HxA (2) ; $T$/K = 298.15

| | | | | | | | |
|---|---|---|---|---|---|---|---|
| 0.0000 | 0.0000 | 3.904 | | 0.6097 | 0.4686 | 10.649 | − 1.008 |
| 0.0708 | 0.0412 | 4.336 | − 0.250 | 0.6977 | 0.5658 | 12.426 | − 0.839 |
| 0.1073 | 0.0635 | 4.578 | − 0.377 | 0.8044 | 0.6989 | 14.948 | − 0.519 |
| 0.1470 | 0.0887 | 4.857 | − 0.515 | 0.8406 | 0.7486 | 15.885 | − 0.405 |
| 0.1935 | 0.1193 | 5.222 | − 0.656 | 0.8989 | 0.8339 | 17.478 | − 0.223 |
| 0.3070 | 0.2001 | 6.286 | − 0.929 | 0.9504 | 0.9154 | 18.955 | − 0.094 |
| 0.3941 | 0.2686 | 7.279 | − 1.069 | 1.0000 | 1.0000 | 20.449 | |
| 0.5056 | 0.3660 | 8.850 | − 1.109 | | | | |

1-propanol (1) + HxA (2) ; $T$/K = 303.15

| | | | | | | | |
|---|---|---|---|---|---|---|---|
| 0.0000 | 0.0000 | 3.846 | | 0.6097 | 0.4685 | 10.324 | − 0.989 |
| 0.0708 | 0.0412 | 4.263 | − 0.240 | 0.6977 | 0.5656 | 12.034 | − 0.827 |
| 0.1073 | 0.0635 | 4.499 | − 0.359 | 0.8044 | 0.6988 | 14.463 | − 0.520 |
| 0.1470 | 0.0886 | 4.770 | − 0.488 | 0.8406 | 0.7484 | 15.354 | − 0.420 |
| 0.1935 | 0.1192 | 5.123 | − 0.623 | 0.8989 | 0.8338 | 16.902 | − 0.233 |
| 0.3070 | 0.2000 | 6.148 | − 0.886 | 0.9504 | 0.9153 | 18.330 | − 0.104 |
| 0.3941 | 0.2685 | 7.102 | − 1.023 | 1.0000 | 1.0000 | 19.784 | |
| 0.5056 | 0.3659 | 8.605 | − 1.073 | | | | |

1-butanol (1) + HxA (2) ; $T$/K = 293.15

| | | | | | | | |
|---|---|---|---|---|---|---|---|
| 0.0000 | 0.0000 | 3.965 | | 0.5930 | 0.5018 | 9.623 | − 1.481 |
| 0.0510 | 0.0358 | 4.258 | − 0.216 | 0.6983 | 0.6154 | 11.376 | − 1.344 |
| 0.0969 | 0.0691 | 4.538 | − 0.410 | 0.8047 | 0.7402 | 13.472 | − 1.024 |
| 0.1402 | 0.1013 | 4.828 | − 0.578 | 0.8525 | 0.7998 | 14.558 | − 0.786 |
| 0.2052 | 0.1515 | 5.294 | − 0.826 | 0.9035 | 0.8662 | 15.741 | − 0.547 |
| 0.3095 | 0.2366 | 6.164 | − 1.167 | 0.9460 | 0.9237 | 16.795 | − 0.311 |
| 0.4059 | 0.3208 | 7.139 | − 1.390 | 1.0000 | 1.0000 | 18.192 | |
| 0.5054 | 0.4140 | 8.352 | − 1.503 | | | | |

1-butanol (1) + HxA (2) ; $T$/K = 298.15

| | | | | | | | |
|---|---|---|---|---|---|---|---|
| 0.0000 | 0.0000 | 3.904 | | 0.5930 | 0.5016 | 9.333 | − 1.413 |
| 0.0510 | 0.0358 | 4.186 | − 0.206 | 0.6983 | 0.6152 | 11.005 | − 1.291 |
| 0.0969 | 0.0690 | 4.460 | − 0.385 | 0.8047 | 0.7400 | 13.003 | − 0.995 |
| 0.1402 | 0.1012 | 4.739 | − 0.545 | 0.8525 | 0.7997 | 14.054 | − 0.759 |
| 0.2052 | 0.1514 | 5.190 | − 0.779 | 0.9035 | 0.8661 | 15.182 | − 0.536 |
| 0.3095 | 0.2364 | 6.028 | − 1.101 | 0.9460 | 0.9237 | 16.197 | − 0.307 |
| 0.4059 | 0.3206 | 6.965 | − 1.312 | 1.0000 | 1.0000 | 17.545 | |
| 0.5054 | 0.4138 | 8.121 | − 1.428 | | | | |

1-butanol (1) + HxA (2) ; $T$/K = 303.15

| | | | | | | | |
|---|---|---|---|---|---|---|---|
| 0.0000 | 0.0000 | 3.845 | | 0.5930 | 0.5014 | 9.058 | − 1.349 |
| 0.0510 | 0.0358 | 4.120 | − 0.194 | 0.6983 | 0.6150 | 10.656 | − 1.238 |
| 0.0969 | 0.0689 | 4.385 | − 0.362 | 0.8047 | 0.7398 | 12.566 | − 0.962 |
| 0.1402 | 0.1012 | 4.656 | − 0.514 | 0.8525 | 0.7996 | 13.565 | − 0.745 |
| 0.2052 | 0.1512 | 5.094 | − 0.730 | 0.9035 | 0.8660 | 14.656 | − 0.523 |
| 0.3095 | 0.2363 | 5.898 | − 1.040 | 0.9460 | 0.9236 | 15.632 | − 0.301 |
| 0.4059 | 0.3204 | 6.797 | − 1.241 | 1.0000 | 1.0000 | 16.933 | |
| 0.5054 | 0.4136 | 7.905 | − 1.353 | | | | |
| | | 1-pentanol (1) + HxA (2) ; $T$/K = 293.15 | | | | | |
| 0.0000 | 0.0000 | 3.967 | | 0.6013 | 0.5521 | 8.827 | − 1.618 |
| 0.0520 | 0.0429 | 4.248 | − 0.222 | 0.6916 | 0.6470 | 10.025 | − 1.534 |
| 0.1079 | 0.0900 | 4.563 | − 0.460 | 0.7952 | 0.7604 | 11.649 | − 1.241 |
| 0.1601 | 0.1348 | 4.876 | − 0.673 | 0.8396 | 0.8106 | 12.419 | − 1.060 |
| 0.2014 | 0.1709 | 5.146 | − 0.826 | 0.9006 | 0.8810 | 13.582 | − 0.723 |
| 0.3100 | 0.2686 | 5.924 | − 1.195 | 0.9431 | 0.9313 | 14.457 | − 0.438 |
| 0.3982 | 0.3510 | 6.653 | − 1.433 | 1.0000 | 1.0000 | 15.701 | |
| 0.5031 | 0.4528 | 7.689 | − 1.591 | | | | |
| | | 1-pentanol (1) + HxA (2) ; $T$/K = 298.15 | | | | | |
| 0.0000 | 0.0000 | 3.905 | | 0.6013 | 0.5519 | 8.567 | − 1.518 |
| 0.0520 | 0.0429 | 4.176 | − 0.209 | 0.6916 | 0.6468 | 9.702 | − 1.445 |
| 0.1079 | 0.0899 | 4.485 | − 0.427 | 0.7952 | 0.7602 | 11.246 | − 1.171 |
| 0.1601 | 0.1347 | 4.785 | − 0.628 | 0.8396 | 0.8104 | 11.973 | − 1.006 |
| 0.2014 | 0.1708 | 5.047 | − 0.770 | 0.9006 | 0.8809 | 13.084 | − 0.684 |
| 0.3100 | 0.2684 | 5.796 | − 1.114 | 0.9431 | 0.9312 | 13.913 | − 0.419 |
| 0.3982 | 0.3508 | 6.494 | − 1.339 | 1.0000 | 1.0000 | 15.102 | |
| 0.5031 | 0.4526 | 7.486 | − 1.487 | | | | |
| | | 1-pentanol (1) + HxA (2) ; $T$/K = 303.15 | | | | | |
| 0.0000 | 0.0000 | 3.846 | | 0.6013 | 0.5516 | 8.319 | − 1.424 |
| 0.0520 | 0.0428 | 4.110 | − 0.194 | 0.6916 | 0.6465 | 9.397 | − 1.360 |
| 0.1079 | 0.0898 | 4.408 | − 0.398 | 0.7952 | 0.7600 | 10.864 | − 1.106 |
| 0.1601 | 0.1345 | 4.701 | − 0.583 | 0.8396 | 0.8102 | 11.555 | − 0.952 |
| 0.2014 | 0.1706 | 4.957 | − 0.713 | 0.9006 | 0.8808 | 12.610 | − 0.652 |
| 0.3100 | 0.2682 | 5.675 | − 1.038 | 0.9431 | 0.9311 | 13.397 | − 0.402 |
| 0.3982 | 0.3505 | 6.348 | − 1.245 | 1.0000 | 1.0000 | 14.536 | |
| 0.5031 | 0.4523 | 7.298 | − 1.383 | | | | |
| | | 1-heptanol (1) + HxA (2) ; $T$/K = 293.15 | | | | | |
| 0.0000 | 0.0000 | 3.963 | | 0.6036 | 0.6191 | 7.425 | − 1.525 |
| 0.0503 | 0.0535 | 4.202 | − 0.192 | 0.7001 | 0.7136 | 8.239 | − 1.473 |
| 0.0950 | 0.1008 | 4.414 | − 0.361 | 0.8007 | 0.8109 | 9.272 | − 1.224 |
| 0.1606 | 0.1696 | 4.731 | − 0.598 | 0.8524 | 0.8604 | 9.877 | − 1.017 |
| 0.2073 | 0.2182 | 4.970 | − 0.751 | 0.8904 | 0.8966 | 10.381 | − 0.805 |
| 0.3067 | 0.3207 | 5.497 | − 1.050 | 0.9402 | 0.9438 | 11.091 | − 0.475 |
| 0.4038 | 0.4196 | 6.064 | − 1.279 | 1.0000 | 1.0000 | 12.019 | |
| 0.4991 | 0.5154 | 6.671 | – 1.444 | | | | |
| | | 1-heptanol (1) + HxA (2) ; $T$/K = 298.15 | | | | | |
| 0.0000 | 0.0000 | 3.903 | | 0.6036 | 0.6188 | 7.224 | – 1.382 |
| 0.0503 | 0.0534 | 4.133 | – 0.176 | 0.7001 | 0.7133 | 7.981 | – 1.344 |
| 0.0950 | 0.1006 | 4.339 | – 0.329 | 0.8007 | 0.8107 | 8.951 | – 1.114 |
| 0.1606 | 0.1694 | 4.644 | – 0.547 | 0.8524 | 0.8603 | 9.521 | – 0.921 |
| 0.2073 | 0.2180 | 4.882 | – 0.678 | 0.8904 | 0.8965 | 9.983 | – 0.734 |
| 0.3067 | 0.3205 | 5.392 | – 0.947 | 0.9402 | 0.9437 | 10.642 | – 0.434 |
| 0.4038 | 0.4193 | 5.934 | – 1.156 | 1.0000 | 1.0000 | 11.504 | |
| 0.4991 | 0.5151 | 6.508 | – 1.310 | | | | |
| | | 1-heptanol (1) + HxA (2) ; $T$/K = 303.15 | | | | | |
| 0.0000 | 0.0000 | 3.848 | | 0.6036 | 0.6185 | 7.035 | – 1.245 |
| 0.0503 | 0.0534 | 4.070 | – 0.161 | 0.7001 | 0.7131 | 7.746 | – 1.212 |
| 0.0950 | 0.1005 | 4.271 | – 0.297 | 0.8007 | 0.8105 | 8.651 | – 1.005 |
| 0.1606 | 0.1692 | 4.567 | – 0.493 | 0.8524 | 0.8601 | 9.177 | – 0.834 |
| 0.2073 | 0.2178 | 4.796 | – 0.613 | 0.8904 | 0.8964 | 9.607 | – 0.665 |
| 0.3067 | 0.3202 | 5.291 | – 0.852 | 0.9402 | 0.9436 | 10.222 | – 0.388 |
| 0.4038 | 0.4190 | 5.811 | – 1.040 | 1.0000 | 1.0000 | 11.014 | |
| 0.4991 | 0.5147 | 6.358 | – 1.178 | | | | |

[a]The standard uncertainties are: $u(T) = 0.02$ K; $u(p) = 1$ kPa; $u(\nu) = 20$ Hz; $u(x_1) = 0.0010$; $u(\phi_1) = 0.004$. The relative standard uncertainty is: $u_r(\varepsilon_r) = 0.003$; and the relative combined expanded uncertainty (0.95 level of confidence) is $U_{rc}(\varepsilon_r^E) = 0.03$.

Table 4

Volume fractions of 1-alkanol, $\phi_1$, refractive indices, $n_D$, and excess refractive indices, $n_D^E$, of 1-alkanol (1) + HxA (2) mixtures as functions of the mole fraction of the 1-alkanol, $x_1$, at temperature $T$ and pressure $p = 0.1$ MPa. [a]

| $x_1$ | $\phi_1$ | $n_D$ | $10^5 n_D^E$ | $x_1$ | $\phi_1$ | $n_D$ | $10^5 n_D^E$ |
|---|---|---|---|---|---|---|---|
| methanol (1) + HxA (2) ; $T/K = 293.15$ ||||||||
| 0.0000 | 0.0000 | 1.41808 |     | 0.6997 | 0.4160 | 1.38836 | 679 |
| 0.0534 | 0.0170 | 1.41741 | 80  | 0.7995 | 0.5494 | 1.37505 | 540 |
| 0.0942 | 0.0308 | 1.41681 | 140 | 0.8475 | 0.6295 | 1.36685 | 440 |
| 0.1871 | 0.0657 | 1.41513 | 275 | 0.8978 | 0.7287 | 1.35657 | 309 |
| 0.2977 | 0.1147 | 1.41250 | 439 | 0.9496 | 0.8521 | 1.34385 | 162 |
| 0.4039 | 0.1716 | 1.40889 | 576 | 0.9834 | 0.9477 | 1.33399 | 54  |
| 0.4917 | 0.2283 | 1.40480 | 664 | 1.0000 | 1.0000 | 1.32862 |     |
| 0.5977 | 0.3124 | 1.39790 | 715 |        |        |         |     |
| methanol (1) + HxA (2) ; $T/K = 298.15$ ||||||||
| 0.0000 | 0.0000 | 1.41563 |     | 0.6997 | 0.4161 | 1.38615 | 689 |
| 0.0534 | 0.0170 | 1.41499 | 83  | 0.7995 | 0.5495 | 1.37296 | 557 |
| 0.0942 | 0.0308 | 1.41442 | 145 | 0.8475 | 0.6296 | 1.36479 | 457 |
| 0.1871 | 0.0658 | 1.41282 | 288 | 0.8978 | 0.7288 | 1.35460 | 332 |
| 0.2977 | 0.1148 | 1.41017 | 448 | 0.9496 | 0.8521 | 1.34182 | 173 |
| 0.4039 | 0.1717 | 1.40657 | 583 | 0.9834 | 0.9477 | 1.33194 | 59  |
| 0.4917 | 0.2283 | 1.40243 | 664 | 1.0000 | 1.0000 | 1.32654 |     |
| 0.5977 | 0.3125 | 1.39550 | 710 |        |        |         |     |
| methanol (1) + HxA (2) ; $T/K = 303.15$ ||||||||
| 0.0000 | 0.0000 | 1.41321 |     | 0.6997 | 0.4162 | 1.38372 | 678 |
| 0.0534 | 0.0170 | 1.41261 | 86  | 0.7995 | 0.5496 | 1.37067 | 556 |
| 0.0942 | 0.0308 | 1.41204 | 148 | 0.8475 | 0.6297 | 1.36257 | 461 |
| 0.1871 | 0.0658 | 1.41040 | 286 | 0.8978 | 0.7289 | 1.35240 | 335 |
| 0.2977 | 0.1148 | 1.40773 | 443 | 0.9496 | 0.8522 | 1.33968 | 179 |
| 0.4039 | 0.1717 | 1.40412 | 576 | 0.9834 | 0.9477 | 1.32980 | 62  |
| 0.4917 | 0.2284 | 1.40000 | 658 | 1.0000 | 1.0000 | 1.32439 |     |
| 0.5977 | 0.3125 | 1.39316 | 709 |        |        |         |     |
| 1-propanol (1) + HxA (2) ; $T/K = 293.15$ ||||||||
| 0.0000 | 0.0000 | 1.41808 |     | 0.6040 | 0.4628 | 1.40784 | 491 |
| 0.0520 | 0.0301 | 1.41775 | 65  | 0.6965 | 0.5645 | 1.40408 | 450 |
| 0.1015 | 0.0600 | 1.41735 | 122 | 0.8013 | 0.6949 | 1.39877 | 350 |
| 0.1475 | 0.0890 | 1.41693 | 175 | 0.8505 | 0.7627 | 1.39588 | 285 |
| 0.2108 | 0.1311 | 1.41627 | 246 | 0.8953 | 0.8285 | 1.39295 | 211 |
| 0.2983 | 0.1936 | 1.41509 | 333 | 0.9486 | 0.9125 | 1.38914 | 109 |
| 0.3967 | 0.2708 | 1.41351 | 427 | 1.0000 | 1.0000 | 1.38514 | |
| 0.4998 | 0.3608 | 1.41113 | 485 | | | | |
| | | 1-propanol (1) + HxA (2) ; $T$/K = 298.15 | | | | | |
| 0.0000 | 0.0000 | 1.41563 | | 0.6040 | 0.4627 | 1.40551 | 487 |
| 0.0520 | 0.0300 | 1.41529 | 63 | 0.6965 | 0.5644 | 1.40199 | 466 |
| 0.1015 | 0.0600 | 1.41492 | 122 | 0.8013 | 0.6948 | 1.39675 | 368 |
| 0.1475 | 0.0890 | 1.41452 | 176 | 0.8505 | 0.7626 | 1.39373 | 288 |
| 0.2108 | 0.1310 | 1.41393 | 253 | 0.8953 | 0.8284 | 1.39085 | 216 |
| 0.2983 | 0.1935 | 1.41276 | 338 | 0.9486 | 0.9124 | 1.38704 | 111 |
| 0.3967 | 0.2707 | 1.41120 | 432 | 1.0000 | 1.0000 | 1.38306 | |
| 0.4998 | 0.3607 | 1.40876 | 480 | | | | |
| | | 1-propanol (1) + HxA (2) ; $T$/K = 303.15 | | | | | |
| 0.0000 | 0.0000 | 1.41321 | | 0.6040 | 0.4625 | 1.40323 | 482 |
| 0.0520 | 0.0300 | 1.41292 | 67 | 0.6965 | 0.5642 | 1.39964 | 450 |
| 0.1015 | 0.0599 | 1.41258 | 128 | 0.8013 | 0.6947 | 1.39454 | 361 |
| 0.1475 | 0.0889 | 1.41223 | 185 | 0.8505 | 0.7625 | 1.39168 | 295 |
| 0.2108 | 0.1310 | 1.41160 | 257 | 0.8953 | 0.8283 | 1.38884 | 224 |
| 0.2983 | 0.1934 | 1.41049 | 345 | 0.9486 | 0.9124 | 1.38505 | 118 |
| 0.3967 | 0.2706 | 1.40878 | 421 | 1.0000 | 1.0000 | 1.38102 | |
| 0.4998 | 0.3605 | 1.40642 | 473 | | | | |
| | | 1-butanol (1) + HxA (2) ; $T$/K = 293.15 | | | | | |
| 0.0000 | 0.0000 | 1.41810 | | 0.6004 | 0.5095 | 1.41298 | 442 |
| 0.0552 | 0.0388 | 1.41812 | 74 | 0.6977 | 0.6148 | 1.41065 | 407 |
| 0.0896 | 0.0637 | 1.41800 | 109 | 0.7982 | 0.7322 | 1.40746 | 309 |
| 0.1588 | 0.1155 | 1.41781 | 187 | 0.8451 | 0.7905 | 1.40577 | 250 |
| 0.1967 | 0.1448 | 1.41768 | 229 | 0.8996 | 0.8610 | 1.40359 | 165 |
| 0.3035 | 0.2315 | 1.41702 | 325 | 0.9461 | 0.9239 | 1.40166 | 91 |
| 0.4077 | 0.3225 | 1.41603 | 396 | 1.0000 | 1.0000 | 1.39931 | |
| 0.4984 | 0.4072 | 1.41487 | 439 | | | | |
| | | 1-butanol (1) + HxA (2) ; $T$/K = 298.15 | | | | | |
| 0.0000 | 0.0000 | 1.41565 | | 0.6004 | 0.5093 | 1.41079 | 444 |
| 0.0552 | 0.0388 | 1.41564 | 70 | 0.6977 | 0.6146 | 1.40845 | 403 |
| 0.0896 | 0.0637 | 1.41556 | 107 | 0.7982 | 0.7321 | 1.40534 | 308 |
| 0.1588 | 0.1154 | 1.41542 | 187 | 0.8451 | 0.7903 | 1.40368 | 249 |
| 0.1967 | 0.1447 | 1.41529 | 228 | 0.8996 | 0.8609 | 1.40158 | 169 |
| 0.3035 | 0.2314 | 1.41470 | 327 | 0.9461 | 0.9238 | 1.39965 | 92 |
| 0.4077 | 0.3223 | 1.41381 | 404 | 1.0000 | 1.0000 | 1.39733 | |
| 0.4984 | 0.4070 | 1.41267 | 445 | | | | |
| | | 1-butanol (1) + HxA (2) ; $T$/K = 303.15 | | | | | |
| 0.0000 | 0.0000 | 1.41322 | | 0.6004 | 0.5091 | 1.40852 | 440 |
| 0.0552 | 0.0388 | 1.41324 | 71 | 0.6977 | 0.6143 | 1.40621 | 398 |
| 0.0896 | 0.0636 | 1.41319 | 110 | 0.7982 | 0.7319 | 1.40316 | 304 |
| 0.1588 | 0.1153 | 1.41306 | 190 | 0.8451 | 0.7902 | 1.40155 | 248 |
| 0.1967 | 0.1446 | 1.41294 | 230 | 0.8996 | 0.8608 | 1.39949 | 169 |
| 0.3035 | 0.2312 | 1.41240 | 331 | 0.9461 | 0.9238 | 1.39765 | 99 |
| 0.4077 | 0.3221 | 1.41148 | 401 | 1.0000 | 1.0000 | 1.39529 | |
| 0.4984 | 0.4068 | 1.41040 | 445 | | | | |

1-pentanol (1) + HxA (2) ; $T/K = 293.15$

| | | | | | | | |
|---|---|---|---|---|---|---|---|
| 0.0000 | 0.0000 | 1.41813 | | 0.6005 | 0.5513 | 1.41756 | 399 |
| 0.0472 | 0.0389 | 1.41838 | 57 | 0.7122 | 0.6692 | 1.41609 | 350 |
| 0.1005 | 0.0837 | 1.41862 | 118 | 0.7963 | 0.7616 | 1.41463 | 280 |
| 0.1648 | 0.1389 | 1.41884 | 186 | 0.8457 | 0.8175 | 1.41364 | 228 |
| 0.2022 | 0.1716 | 1.41896 | 225 | 0.8982 | 0.8782 | 1.41246 | 160 |
| 0.3102 | 0.2688 | 1.41908 | 317 | 0.9362 | 0.9230 | 1.41153 | 104 |
| 0.3982 | 0.3510 | 1.41895 | 372 | 1.0000 | 1.0000 | 1.40985 | |
| 0.5002 | 0.4500 | 1.41846 | 405 | | | | |

1-pentanol (1) + HxA (2) ; $T/K = 298.15$

| | | | | | | | |
|---|---|---|---|---|---|---|---|
| 0.0000 | 0.0000 | 1.41563 | | 0.6005 | 0.5510 | 1.41543 | 404 |
| 0.0472 | 0.0389 | 1.41592 | 59 | 0.7122 | 0.6689 | 1.41403 | 355 |
| 0.1005 | 0.0836 | 1.41622 | 123 | 0.7963 | 0.7614 | 1.41259 | 282 |
| 0.1648 | 0.1388 | 1.41649 | 193 | 0.8457 | 0.8174 | 1.41164 | 230 |
| 0.2022 | 0.1715 | 1.41660 | 229 | 0.8982 | 0.8781 | 1.41048 | 161 |
| 0.3102 | 0.2686 | 1.41679 | 322 | 0.9362 | 0.9230 | 1.40957 | 105 |
| 0.3982 | 0.3508 | 1.41671 | 378 | 1.0000 | 1.0000 | 1.40793 | |
| 0.5002 | 0.4497 | 1.41623 | 406 | | | | |

1-pentanol (1) + HxA (2) ; $T/K = 303.15$

| | | | | | | | |
|---|---|---|---|---|---|---|---|
| 0.0000 | 0.0000 | 1.41324 | | 0.6005 | 0.5508 | 1.41329 | 409 |
| 0.0472 | 0.0388 | 1.41360 | 64 | 0.7122 | 0.6687 | 1.41193 | 359 |
| 0.1005 | 0.0835 | 1.41392 | 129 | 0.7963 | 0.7612 | 1.41054 | 288 |
| 0.1648 | 0.1386 | 1.41424 | 202 | 0.8457 | 0.8172 | 1.40958 | 234 |
| 0.2022 | 0.1713 | 1.41438 | 239 | 0.8982 | 0.8780 | 1.40844 | 164 |
| 0.3102 | 0.2683 | 1.41458 | 331 | 0.9362 | 0.9229 | 1.40754 | 107 |
| 0.3982 | 0.3505 | 1.41451 | 384 | 1.0000 | 1.0000 | 1.40590 | |
| 0.5002 | 0.4494 | 1.41407 | 412 | | | | |

1-heptanol (1) + HxA (2) ; $T/K = 293.15$

| | | | | | | | |
|---|---|---|---|---|---|---|---|
| 0.0000 | 0.0000 | 1.41807 | | 0.6020 | 0.6175 | 1.42550 | 361 |
| 0.0560 | 0.0596 | 1.41916 | 72 | 0.7029 | 0.7163 | 1.42563 | 313 |
| 0.1018 | 0.1079 | 1.41999 | 125 | 0.7979 | 0.8082 | 1.42545 | 238 |
| 0.1463 | 0.1546 | 1.42077 | 174 | 0.8534 | 0.8614 | 1.42521 | 181 |
| 0.2047 | 0.2155 | 1.42171 | 231 | 0.8986 | 0.9044 | 1.42496 | 130 |
| 0.3044 | 0.3184 | 1.42310 | 306 | 0.9472 | 0.9504 | 1.42464 | 70 |
| 0.4059 | 0.4217 | 1.42423 | 355 | 1.0000 | 1.0000 | 1.42425 | |
| 0.5044 | 0.5207 | 1.42503 | 374 | | | | |
| | | 1-heptanol (1) + HxA (2) ; $T$/K = 298.15 | | | | | |
| 0.0000 | 0.0000 | 1.41563 | | 0.6020 | 0.6172 | 1.42331 | 353 |
| 0.0560 | 0.0595 | 1.41674 | 71 | 0.7029 | 0.7161 | 1.42351 | 306 |
| 0.1018 | 0.1078 | 1.41761 | 125 | 0.7979 | 0.8080 | 1.42339 | 233 |
| 0.1463 | 0.1545 | 1.41840 | 173 | 0.8534 | 0.8612 | 1.42320 | 178 |
| 0.2047 | 0.2153 | 1.41937 | 229 | 0.8986 | 0.9043 | 1.42299 | 128 |
| 0.3044 | 0.3181 | 1.42080 | 303 | 0.9472 | 0.9503 | 1.42271 | 69 |
| 0.4059 | 0.4214 | 1.42197 | 350 | 1.0000 | 1.0000 | 1.42235 | |
| 0.5044 | 0.5204 | 1.42280 | 367 | | | | |
| | | 1-heptanol (1) + HxA (2) ; $T$/K = 303.15 | | | | | |
| 0.0000 | 0.0000 | 1.41321 | | 0.6020 | 0.6169 | 1.42116 | 347 |
| 0.0560 | 0.0594 | 1.41434 | 70 | 0.7029 | 0.7158 | 1.42142 | 301 |
| 0.1018 | 0.1077 | 1.41523 | 124 | 0.7979 | 0.8078 | 1.42136 | 228 |
| 0.1463 | 0.1543 | 1.41604 | 171 | 0.8534 | 0.8611 | 1.42120 | 174 |
| 0.2047 | 0.2151 | 1.41703 | 226 | 0.8986 | 0.9042 | 1.42103 | 125 |
| 0.3044 | 0.3178 | 1.41853 | 301 | 0.9472 | 0.9502 | 1.42079 | 68 |
| 0.4059 | 0.4211 | 1.41974 | 347 | 1.0000 | 1.0000 | 1.42047 | |
| 0.5044 | 0.5200 | 1.42062 | 363 | | | | |

[a]The standard uncertainties are: $u(T) = 0.02$ K; $u(p) = 1$ kPa; $u(x_1) = 0.0008$; $u(\phi_1) = 0.004$, $u(n_D) = 0.00008$. The combined expanded uncertainty (0.95 level of confidence) is $U_{rc}(n_D^E) = 0.0002$.

Table 5

Coefficients $A_i$ and standard deviations, $\sigma(F^E)$ (equation (6)), for the representation of $F^E$ at temperature $T$ and pressure $p = 0.1$ MPa for 1-alkanol (1) + HxA (2) systems by equation (5).

| Property $F^E$ | System | $T$/K | $A_0$ | $A_1$ | $A_2$ | $A_3$ | $A_3$ | $\sigma(F^E)$ |
|---|---|---|---|---|---|---|---|---|
| $\varepsilon_r^E$ | methanol + HxA | 293.15 | 1.12 | 9.1 | 8.4 | 2.2 | | 0.016 |
| | | 298.15 | 0.92 | 8.4 | 8.5 | 3.1 | | 0.017 |
| | | 303.15 | 0.82 | 8.0 | 8.6 | 3.4 | | 0.013 |
| | 1-propanol + HxA | 293.15 | −4.59 | 0.66 | 2.08 | 0.65 | | 0.004 |
| | | 298.15 | −4.45 | 0.44 | 1.92 | 0.8 | | 0.005 |
| | | 303.15 | −4.30 | 0.23 | 1.74 | 0.8 | | 0.005 |
| | 1-butanol + HxA | 293.15 | −6.00 | −1.1 | 0.8 | | | 0.012 |
| | | 298.15 | −5.70 | −1.17 | 0.6 | | | 0.013 |
| | | 303.15 | −5.41 | −1.25 | 0.49 | | | 0.011 |
| | 1-pentanol + HxA | 293.15 | −6.38 | −2.08 | | | | 0.006 |
| | | 298.15 | −5.98 | −2.02 | | | | 0.008 |
| | | 303.15 | −5.55 | −1.98 | −0.24 | | | 0.006 |
| | 1-heptanol + HxA | 293.15 | −5.81 | −2.6 | −0.79 | | | 0.007 |
| | | 298.15 | −5.26 | −2.38 | −0.77 | | | 0.007 |
| | | 303.15 | −4.73 | −2.15 | −0.73 | | | 0.008 |
| $10^5 n_D^E$ | methanol + HxA | 293.15 | 2694 | 1518 | −255 | −670 | | 2 |
| | | 298.15 | 2694 | 1454 | −22 | −427 | | 4 |
| | | 303.15 | 2663 | 1443 | 64 | −385 | | 0.8 |
| | 1-propanol + HxA | 293.15 | 1922 | 614 | −230 | | | 5 |
| | | 298.15 | 1898 | 644 | | | | 8 |
| | | 303.15 | 1892 | 609 | | | | 1 |
| | 1-butanol + HxA | 293.15 | 1755 | 551 | −215 | −406 | | 3 |
| | | 298.15 | 1772 | 491 | −264 | −260 | | 1.5 |
| | | 303.15 | 1761 | 363 | −218 | | | 4 |
| | 1-pentanol + HxA | 293.15 | 1615 | 283 | −146 | | | 0.9 |
| | | 298.15 | 1631 | 268 | −119 | | | 1.5 |
| | | 303.15 | 1653 | 243 | −62 | | | 1.2 |
| | 1-heptanol + HxA | 293.15 | 1495 | 50 | −147 | | | 1.5 |
| | | 298.15 | 1468 | 33 | −126 | | | 0.9 |
| | | 303.15 | 1451 | 20 | −138 | | | 0.6 |
| $\left(\dfrac{\partial \varepsilon_r^E}{\partial T}\right)_p$ / K$^{-1}$ | methanol + HxA | 298.15 | −0.028 | −0.105 | −0.01 | 0.11 | 0.05 | 0.0007 |
| | 1-propanol + HxA | 298.15 | 0.0284 | −0.038 | −0.032 | | | 0.0003 |
| | 1-butanol + HxA | 298.15 | 0.060 | −0.015 | −0.034 | | | 0.0005 |
| | 1-pentanol + HxA | 298.15 | 0.082 | 0.010 | −0.017 | | | 0.0005 |
| | 1-heptanol + HxA | 298.15 | 0.109 | 0.045 | | | | 0.0007 |

Table 6

Intrinsic dipole moment, $\mu$, and molar dielectric properties of pure liquids at $p = 0.1$ MPa: $\chi_m^*$, molar dielectric susceptibility; $\eta_{m,p}^* = -\left(\partial \chi_m^*/\partial T\right)_p$.

| Compound | $\mu$ / D | $\chi_m^*$/cm³·mol⁻¹ | | | $\eta_{m,p}^*$/cm³·mol⁻¹·K⁻¹ |
|---|---|---|---|---|---|
| | | $T$ = 293.15 K | $T$ = 298.15 K | $T$ = 303.15 K | $T$ = 298.15 K |
| methanol | 1.664 [75] | 1318 | 1287 | 1256 | 6.2 |
| 1-propanol | 1.629 [75] | 1507 | 1462 | 1419 | 8.8 |
| 1-butanol | 1.614 [75] | 1574 | 1521 | 1472 | 10.2 |
| 1-pentanol | 1.598 [75] | 1591 | 1533 | 1478 | 11.3 |
| 1-heptanol | 1.583 [75] | 1557 | 1491 | 1427 | 13.0 |
| 1-decanol | 1.566 [75] | 1413 [a] | 1342 [a] | 1277 [a] | 13.6 [a] |
| HxA | 1.3 [76] | 392 | 387 | 381 | 1.1 |
| c-HxA | 1.26 [77] | 418 [a] | 409 [a] | 401 [a] | 1.7 [a] |
| aniline | 1.51 [40] | 558 [b] | 548 [b] | 538 [b] | 2.0 [b] |

[a] Calculated from data of ref. [19].

[b] Calculated from data of ref. [78].

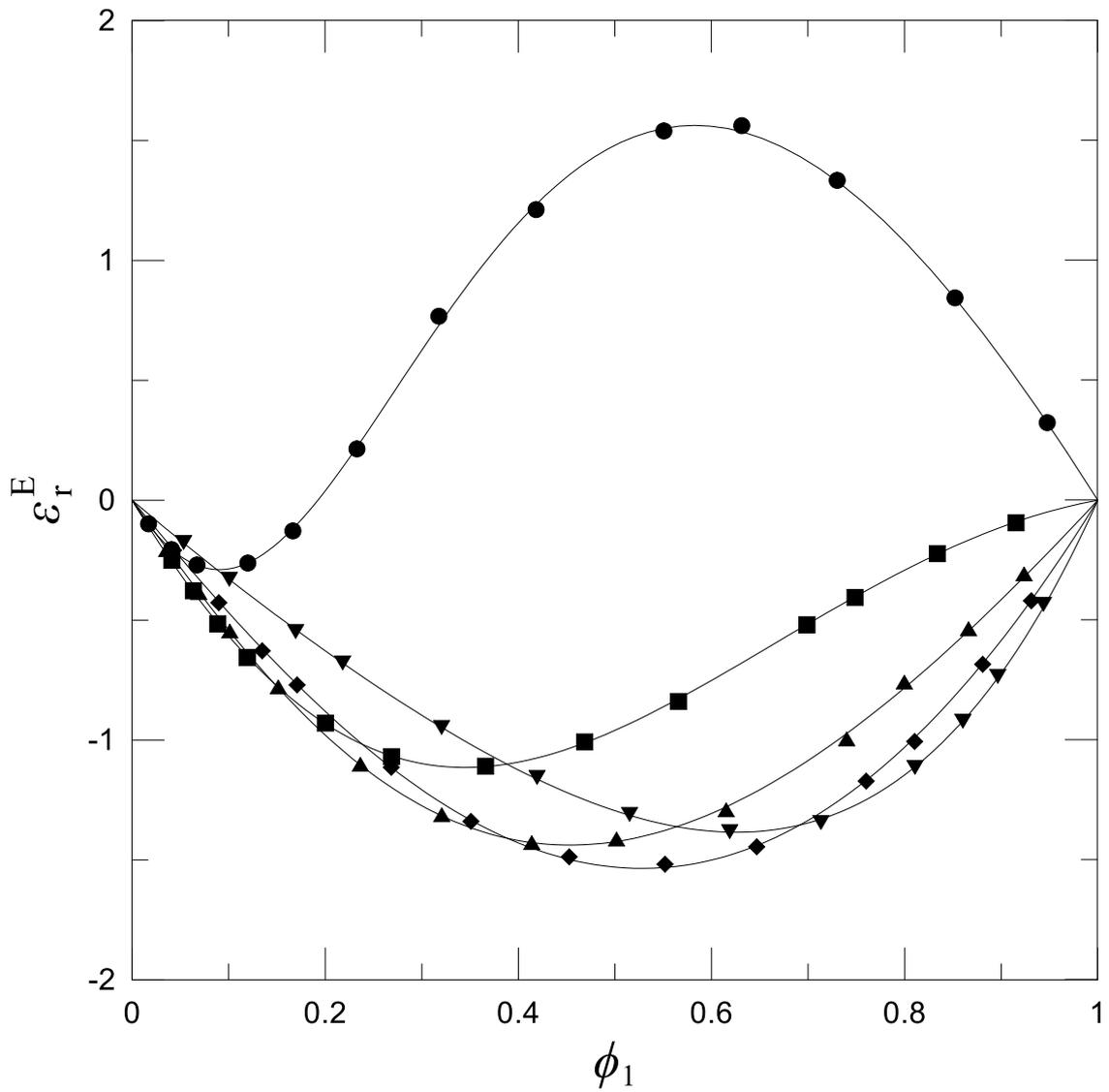

Figure 1

Excess relative permittivities, $\varepsilon_r^E$, of 1-alkanol (1) + HxA (2) systems at 0.1 MPa, 298.15 K and 1 MHz. Full symbols, experimental values (this work): (●), methanol; (■), 1-propanol; (▲), 1-butanol; (♦), 1-pentanol; (▼), 1-heptanol. Solid lines, calculations with equation (5) using the coefficients from Table 5.

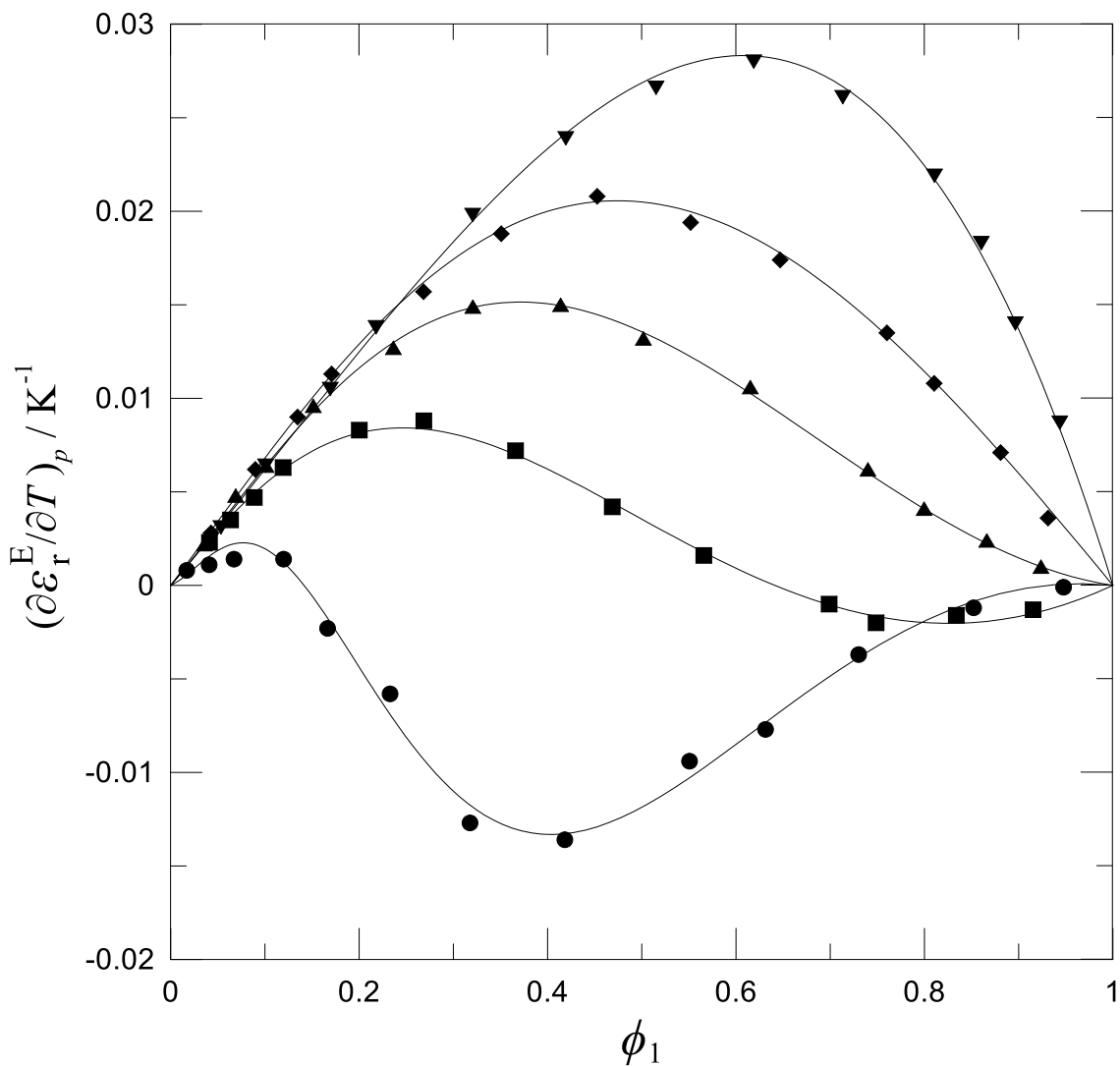

Figure 2

Derivative of the excess relative permittivity of 1-alkanol (1) + HxA (2) systems at 0.1 MPa, 298.15 K and 1 MHz. Full symbols, experimental values (this work): (●), methanol; (■), 1-propanol; (▲), 1-butanol; (♦), 1-pentanol; (▼), 1-heptanol. Solid lines, calculations with equation (5) using the coefficients from Table 5.

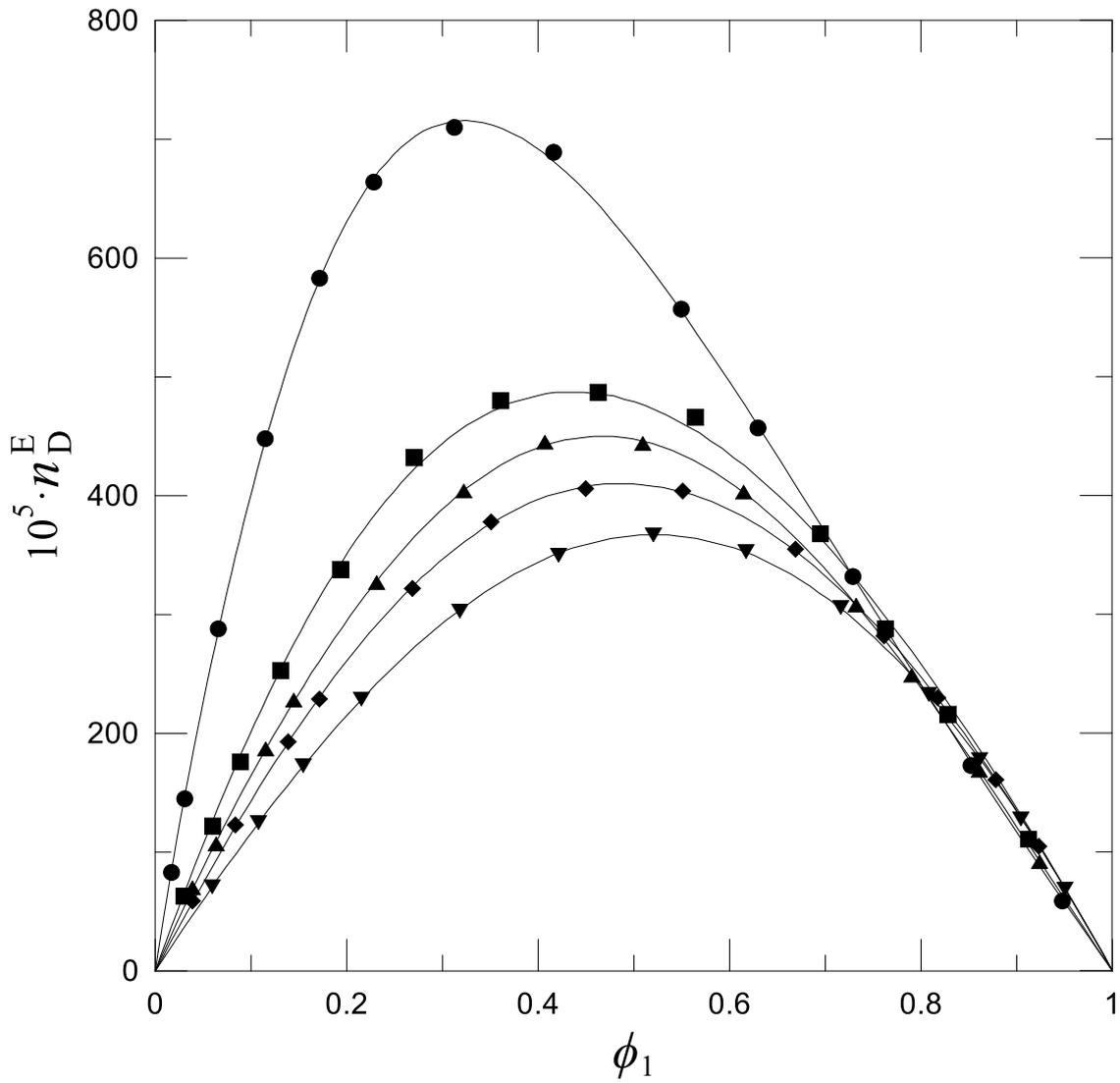

Figure 3

Excess refractive index, $n_D^E$, of 1-alkanol (1) + HxA (2) systems at 0.1 MPa, 298.15 K and 1 MHz. Full symbols, experimental values (this work): (●), methanol; (■), 1-propanol; (▲), 1-butanol; (♦), 1-pentanol; (▼), 1-heptanol. Solid lines, calculations with equation (5) using the coefficients from Table 5.

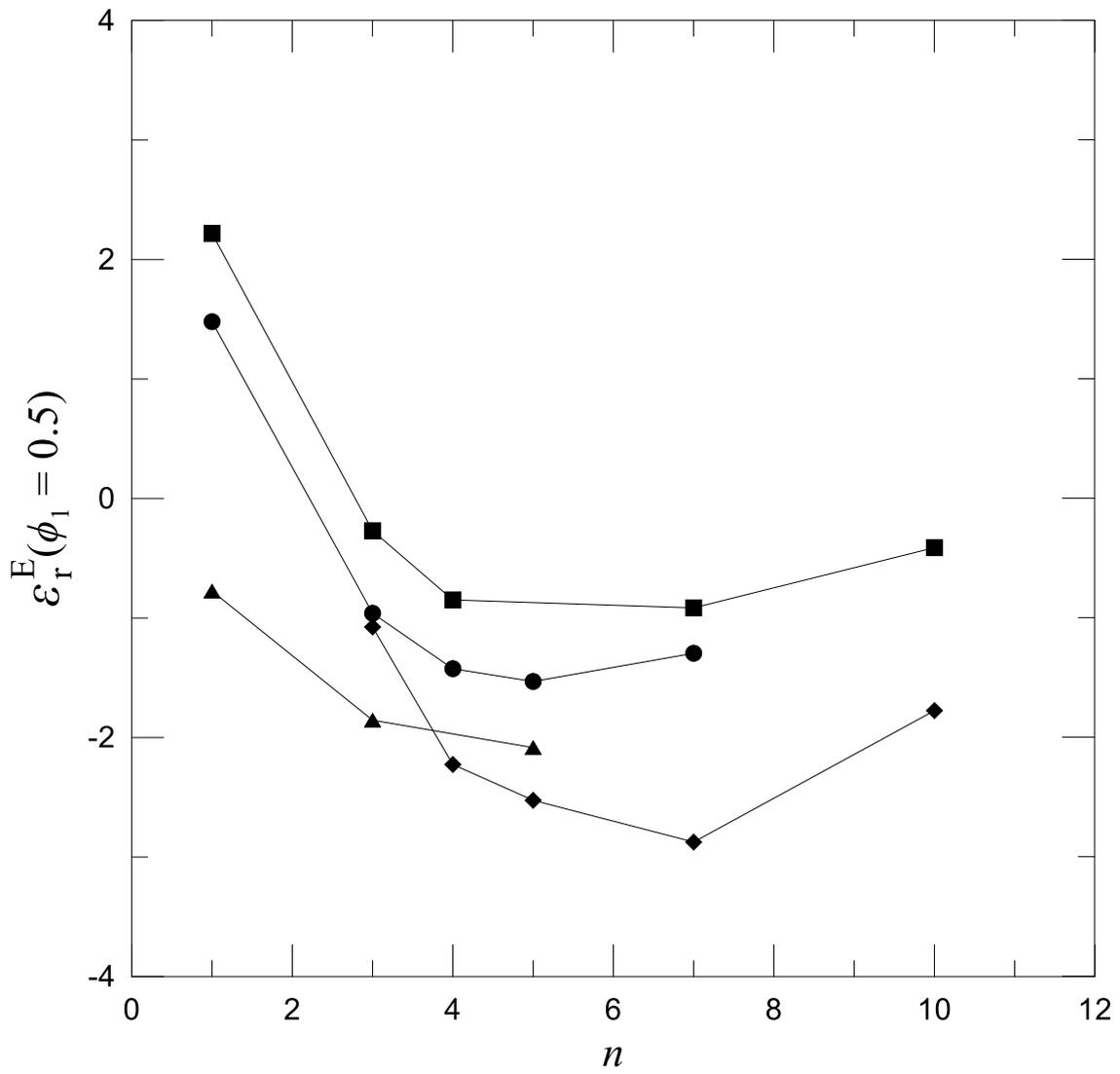

Figure 4

Excess relative permittivities at $\phi_1 = 0.5$ of 1-alkanol (1) + amine (2) or + heptane (2) systems as functions of the number of carbon atoms of the 1-alkanol, at 0.1 MPa, 298.15 K and 1 MHz: (●), HxA (this work); (■), c-HxA [13, 19]; (▲), aniline [42]; (♦), heptane [12, 38-40].

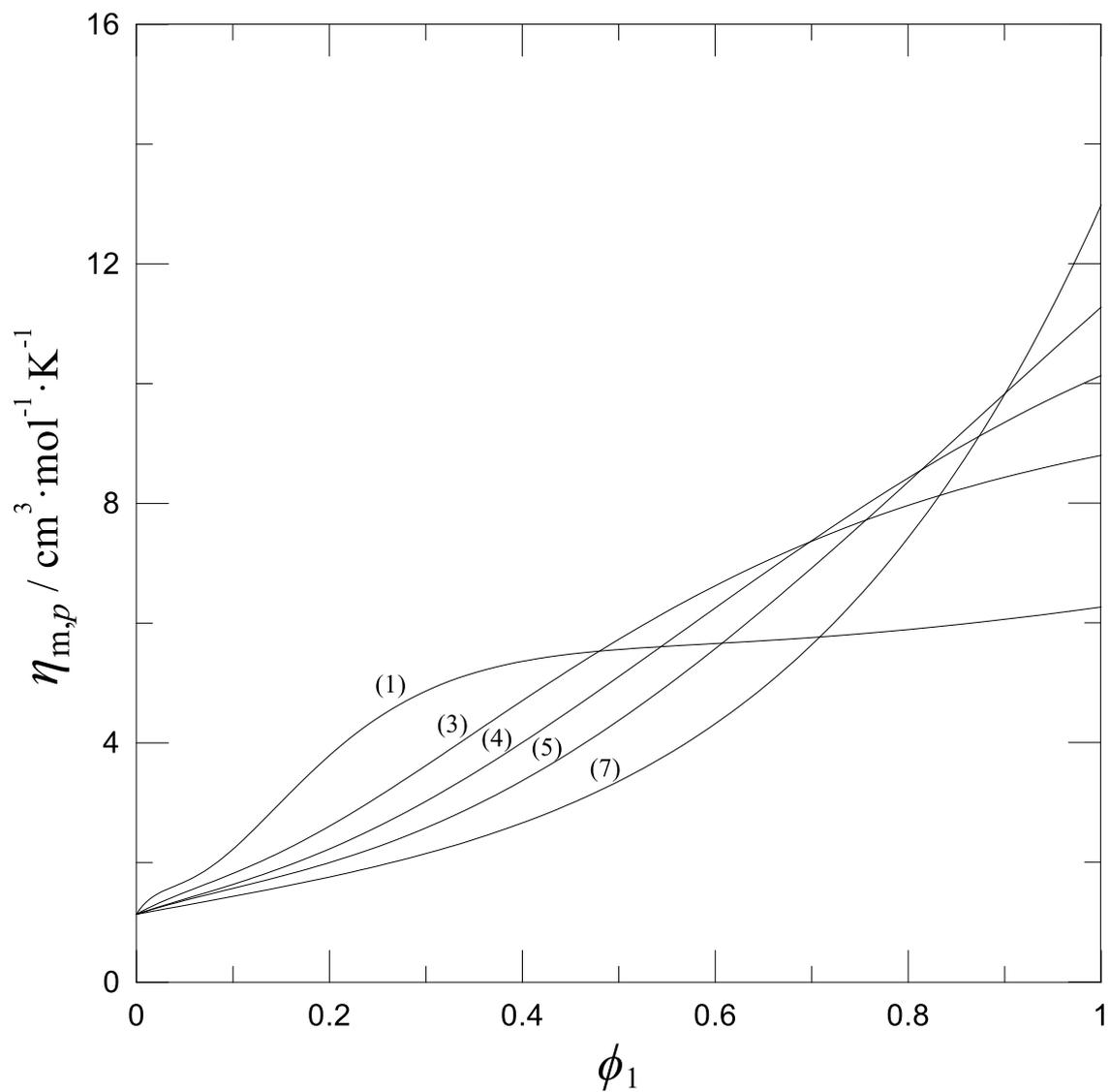

Figure 5

$\eta_{m,p} = -\left(\dfrac{\partial \chi_m}{\partial T}\right)_p$ of 1-alkanol (1) + HxA (2) systems at 0.1 MPa, 298.15 K and 1 MHz.

Numbers in parentheses indicate the number of atoms of the 1-alkanol.

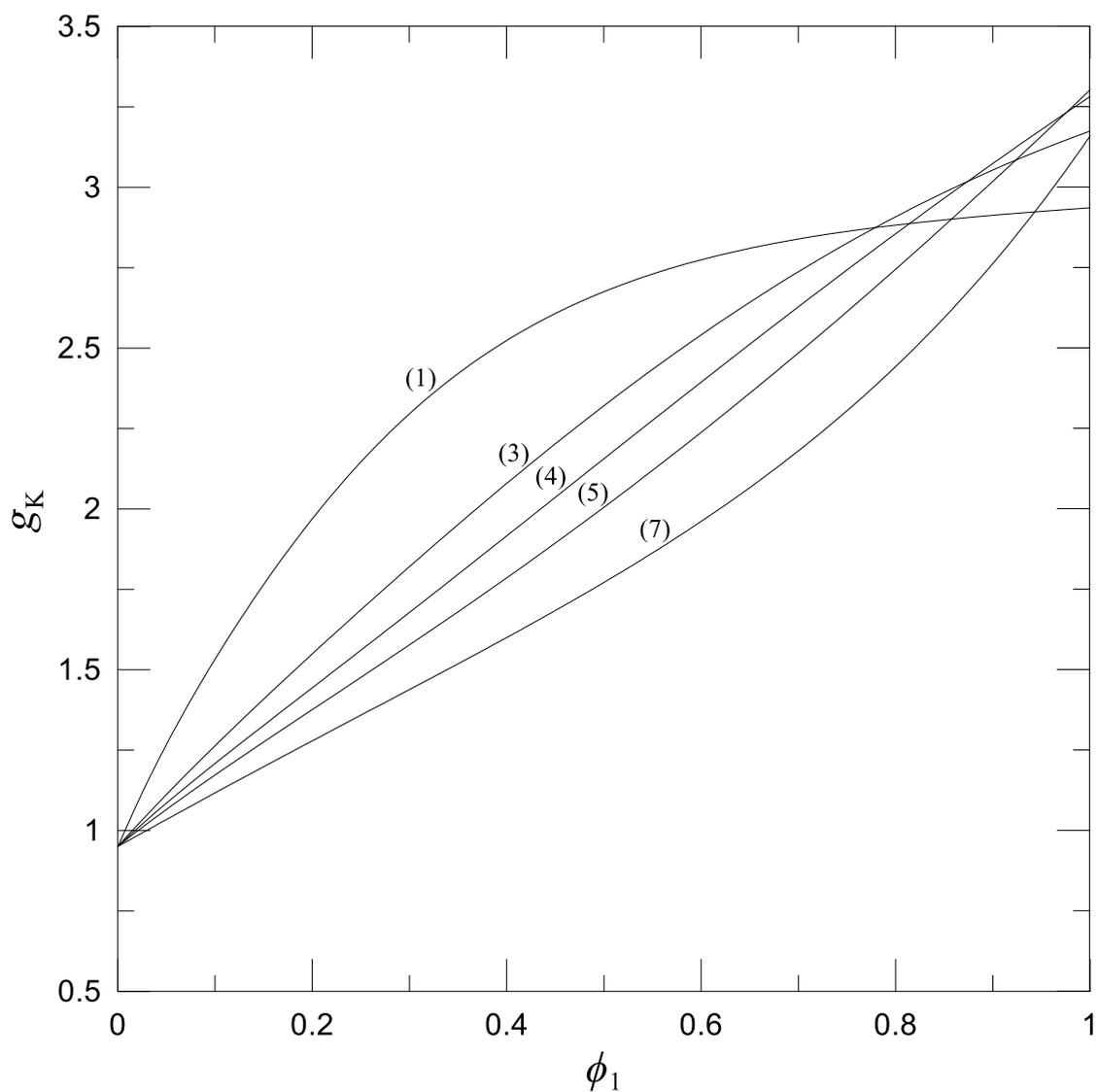

Figure 6

Kirkwood correlation factor, $g_K$, of 1-alkanol (1) + HxA (2) systems at 0.1 MPa, 298.15 K and 1 MHz. Numbers in parentheses indicate the number of atoms of the 1-alkanol.

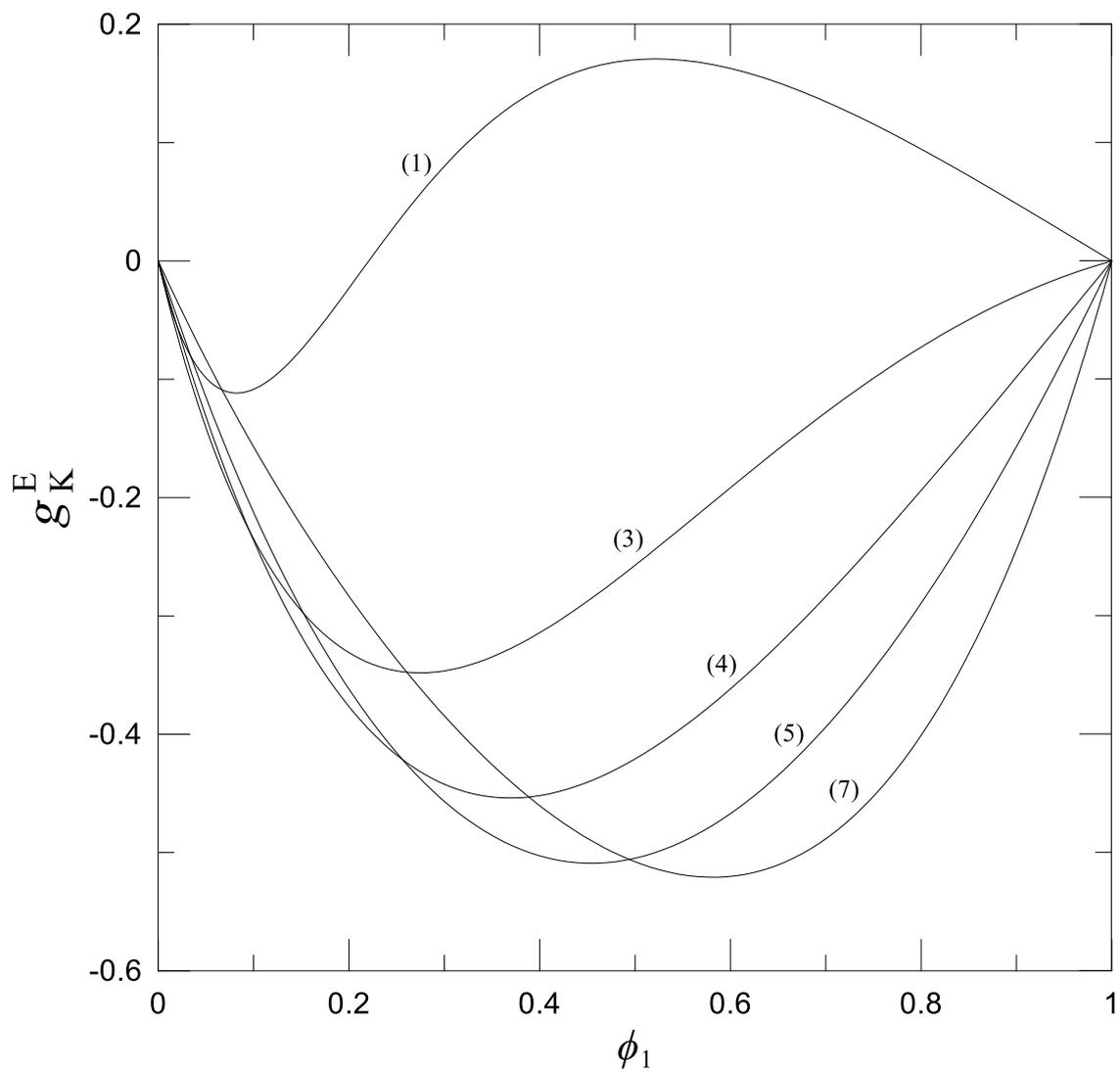

Figure 7

Excess Kirkwood correlation factor, $g_K^E$, of 1-alkanol (1) + HxA (2) systems at 0.1 MPa, 298.15 K and 1 MHz. Numbers in parentheses indicate the number of atoms of the 1-alkanol.

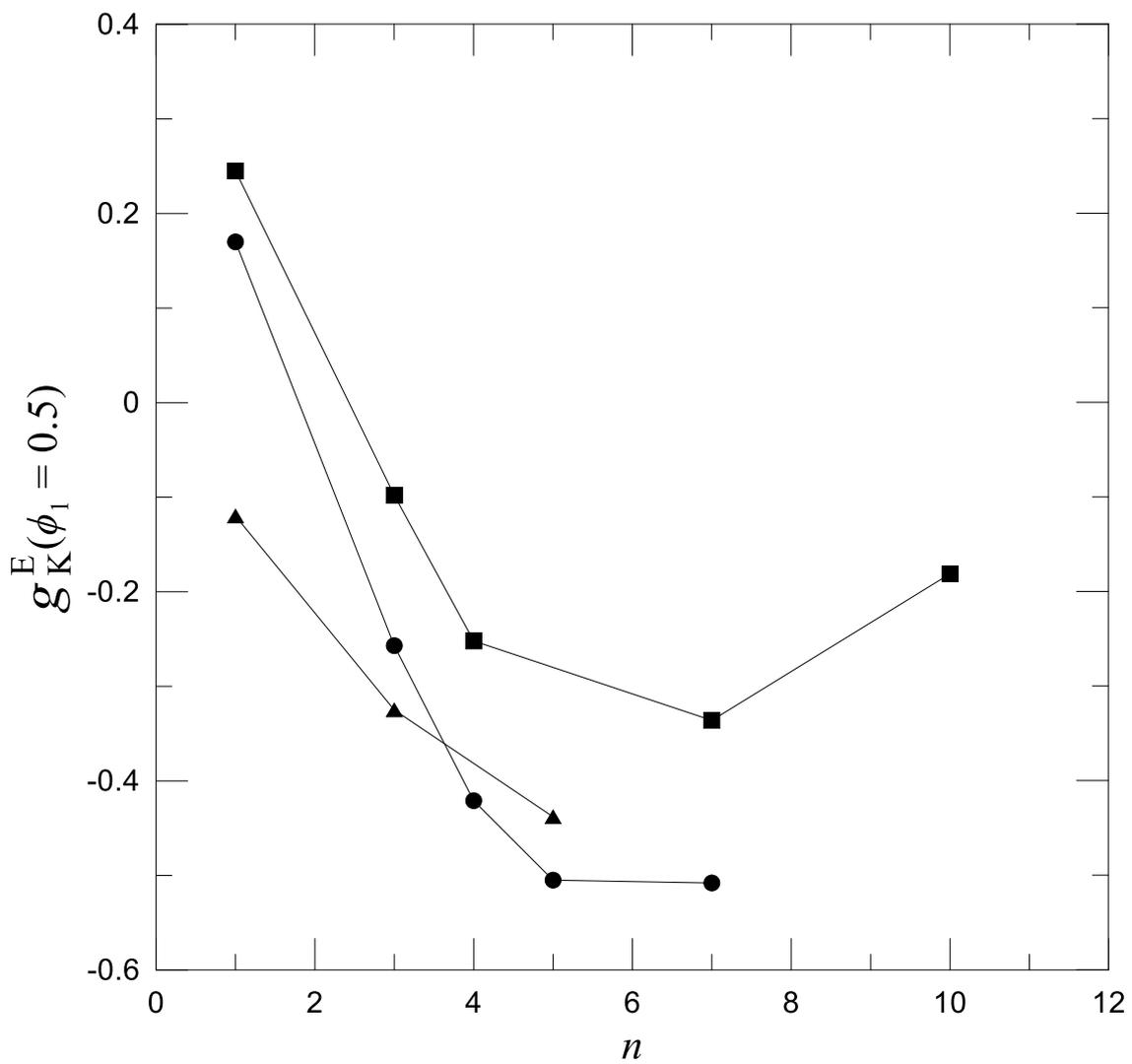

Figure 8

Excess Kirkwood correlation factors at $\phi_1 = 0.5$ of 1-alkanol (1) + amine (2) systems as functions of the number of carbon atoms of the 1-alkanol, at 0.1 MPa, 298.15 K and 1 MHz: (●), HxA (this work); (■), c-HxA [13, 19]; (▲), aniline [42].

**Thermodynamics of mixtures with strongly negative deviations from Raoult's law. XV. Permittivities and refractive indices for 1-alkanol + *n*-hexylamine systems at (293.15-303.15) K. Application of the Kirkwood-Fröhlich model**

**Supplementary material**

Fernando Hevia[(1)], Juan Antonio González[(1)]\*, Ana Cobos[(1)], Isaías García de la Fuente[(1)], Cristina Alonso-Tristán[(2)]

[(1)]G.E.T.E.F., Departamento de Física Aplicada, Facultad de Ciencias, Universidad de Valladolid. Paseo de Belén, 7, 47011 Valladolid, Spain.

[(2)]Unidad de Investigación Consolidada UIC-011, JCyL. Departamento de Ingeniería Electromecánica, Escuela Politécnica Superior, Universidad de Burgos. Avda. Cantabria s/n. 09006, Burgos, Spain.

\*e-mail: jagl@termo.uva.es; Tel: +34-983-423757

Table S1

Derivative of the excess relative permittivity of 1-alkanol (1) + HxA (2) systems at 0.1 MPa, 298.15 K and 1 MHz [a].

| $x_1$ | $\phi_1$ | $(\partial \varepsilon_r^E / \partial T)_p$ / K$^{-1}$ | $x_1$ | $\phi_1$ | $(\partial \varepsilon_r^E / \partial T)_p$ / K$^{-1}$ |
|---|---|---|---|---|---|
| | | methanol (1) + HxA (2) | | | |
| 0.0534 | 0.0170 | 0.0008 | 0.6997 | 0.4183 | –0.0136 |
| 0.0942 | 0.0408 | 0.0011 | 0.7995 | 0.5506 | –0.0094 |
| 0.1871 | 0.0672 | 0.0014 | 0.8475 | 0.6314 | –0.0077 |
| 0.2977 | 0.1199 | 0.0014 | 0.8978 | 0.7301 | –0.0037 |
| 0.4039 | 0.1665 | –0.0023 | 0.9496 | 0.8521 | –0.0012 |
| 0.4917 | 0.2329 | –0.0058 | 0.9834 | 0.9477 | –0.0001 |
| 0.5977 | 0.3177 | –0.0127 | | | |
| | | 1–propanol (1) + HxA (2) | | | |
| 0.0520 | 0.0412 | 0.0023 | 0.6040 | 0.4686 | 0.0042 |
| 0.1015 | 0.0635 | 0.0035 | 0.6965 | 0.5658 | 0.0016 |
| 0.1475 | 0.0887 | 0.0047 | 0.8013 | 0.6989 | –0.0010 |
| 0.2108 | 0.1193 | 0.0063 | 0.8505 | 0.7486 | –0.0020 |
| 0.2983 | 0.2001 | 0.0083 | 0.8953 | 0.8339 | –0.0016 |
| 0.3967 | 0.2686 | 0.0088 | 0.9486 | 0.9154 | –0.0013 |
| 0.4998 | 0.3660 | 0.0072 | | | |
| | | 1–butanol (1) + HxA (2) | | | |
| 0.0552 | 0.0358 | 0.0022 | 0.6004 | 0.5016 | 0.0132 |
| 0.0896 | 0.0690 | 0.0048 | 0.6977 | 0.6152 | 0.0106 |
| 0.1588 | 0.1012 | 0.0064 | 0.7982 | 0.7400 | 0.0062 |
| 0.1967 | 0.1514 | 0.0096 | 0.8451 | 0.7997 | 0.0041 |
| 0.3035 | 0.2364 | 0.0127 | 0.8996 | 0.8661 | 0.0024 |
| 0.4077 | 0.3206 | 0.0149 | 0.9461 | 0.9237 | 0.0010 |
| 0.4984 | 0.4138 | 0.0150 | | | |
| | | 1–pentanol (1) + HxA (2) | | | |
| 0.0472 | 0.0429 | 0.0028 | 0.6005 | 0.5519 | 0.0194 |
| 0.1005 | 0.0899 | 0.0062 | 0.7122 | 0.6468 | 0.0174 |
| 0.1648 | 0.1347 | 0.0090 | 0.7963 | 0.7602 | 0.0135 |
| 0.2022 | 0.1708 | 0.0113 | 0.8457 | 0.8104 | 0.0108 |
| 0.3102 | 0.2684 | 0.0157 | 0.8982 | 0.8809 | 0.0071 |
| 0.3982 | 0.3508 | 0.0188 | 0.9362 | 0.9312 | 0.0036 |
| 0.5002 | 0.4526 | 0.0208 | | | |
| | | 1–heptanol (1) + HxA (2) | | | |
| 0.0560 | 0.0534 | 0.0031 | 0.6020 | 0.6188 | 0.0280 |
| 0.1018 | 0.1006 | 0.0064 | 0.7029 | 0.7133 | 0.0261 |
| 0.1463 | 0.1694 | 0.0105 | 0.7979 | 0.8107 | 0.0219 |
| 0.2047 | 0.2180 | 0.0138 | 0.8534 | 0.8603 | 0.0183 |
| 0.3044 | 0.3205 | 0.0198 | 0.8986 | 0.8965 | 0.014 |
| 0.4059 | 0.4193 | 0.0239 | 0.9472 | 0.9437 | 0.0087 |
| 0.5044 | 0.5151 | 0.0266 | | | |

[a] The standard uncertainties are: $u(T) = 0.02$ K; $u(p) = 1$ kPa; $u(\nu) = 20$ Hz; $u(x_1) = 0.0010$; $u(\phi_1) = 0.004$. The standard uncertainty is: $u\left[\left(\partial \varepsilon_r^E / \partial T\right)_p\right] = 0.0008$.

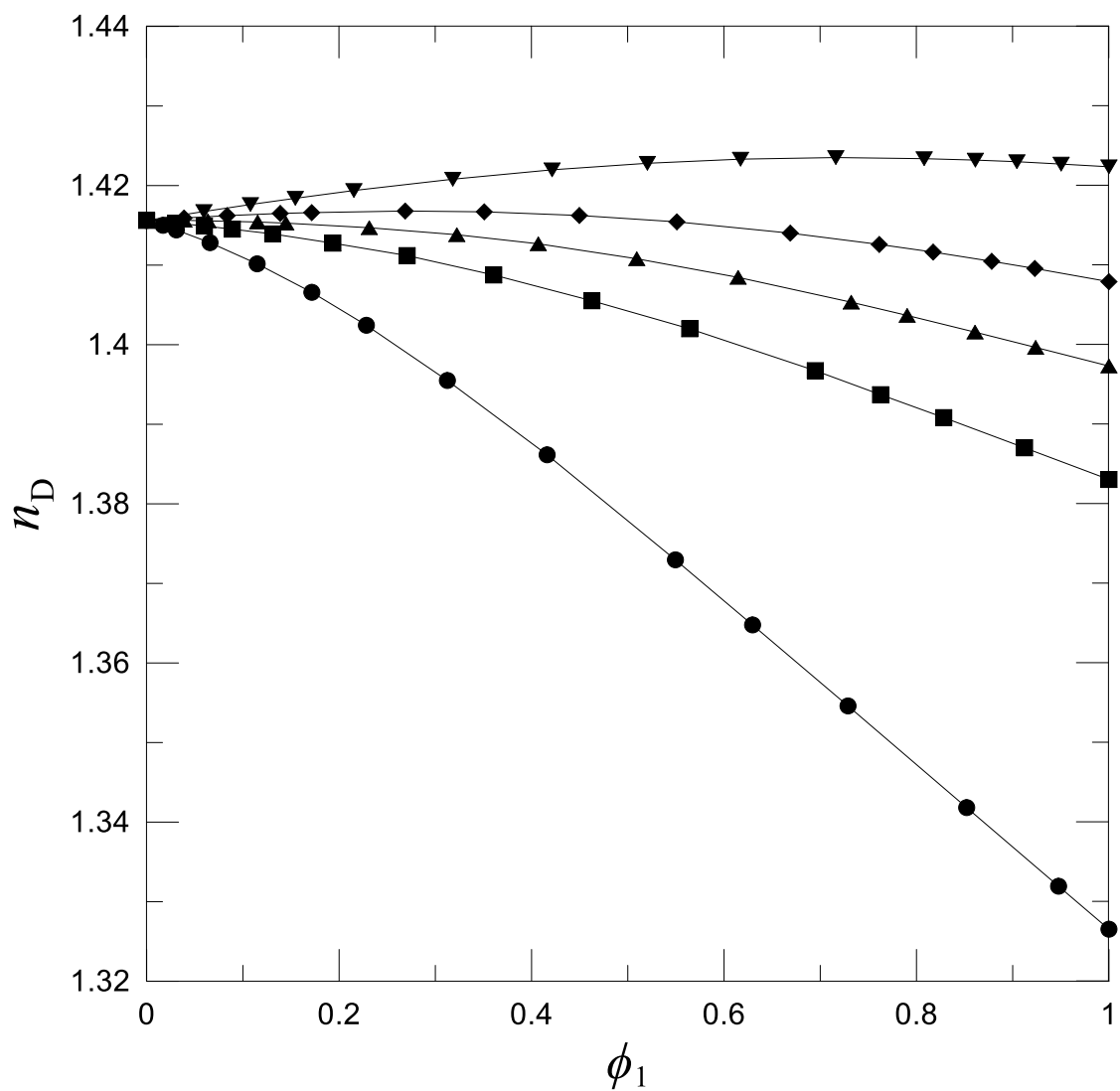

Figure S1

Refractive index at the sodium D line, $n_{\mathrm{D}}$, of 1-alkanol (1) + HxA (2) systems at 0.1 MPa and 298.15 K. Full symbols, experimental values (this work): (●), methanol; (■), 1-propanol; (▲), 1-butanol; (♦), 1-pentanol; (▼), 1-heptanol.

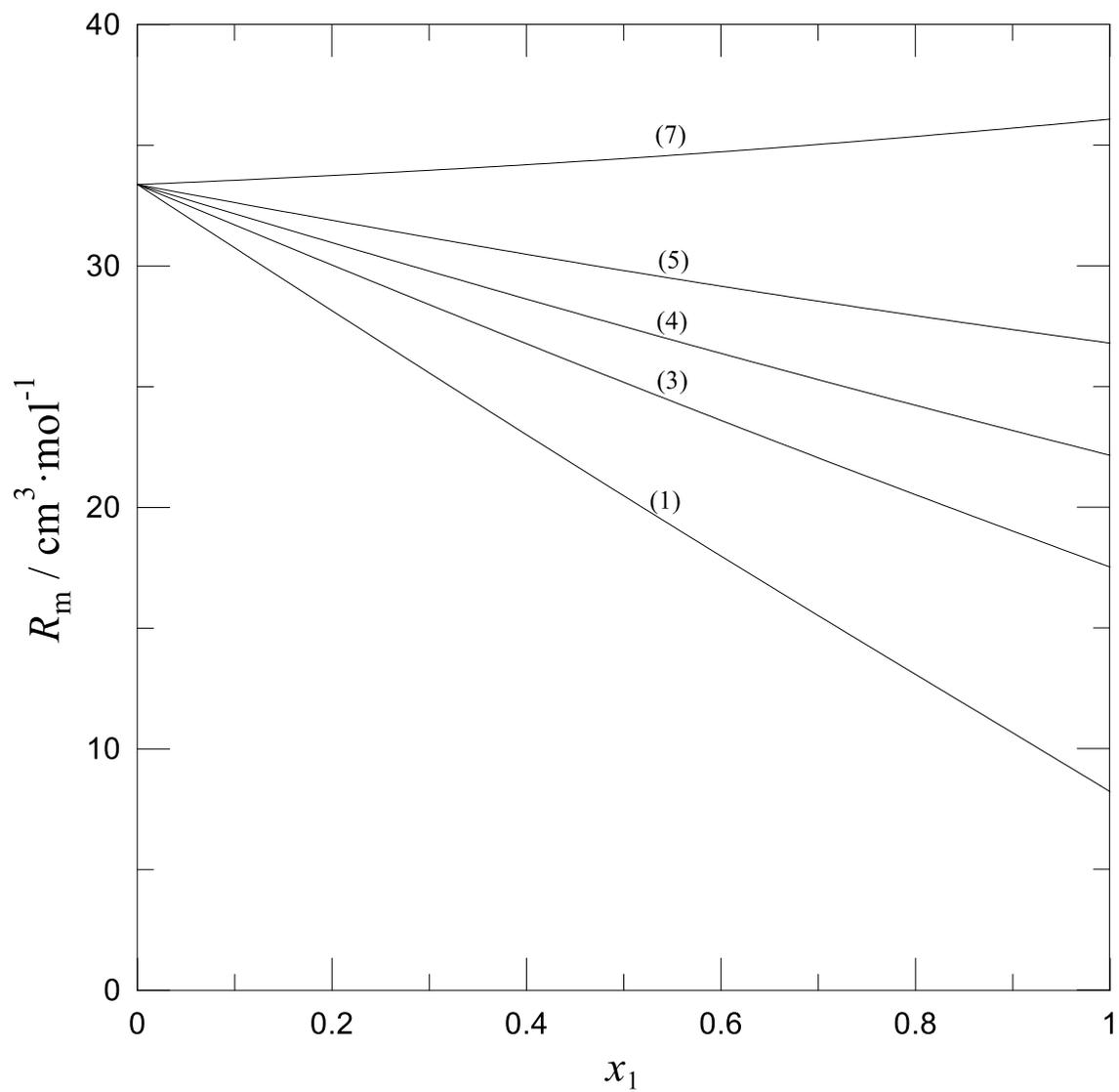

Figure S2

Molar refraction of 1-alkanol (1) + HxA (2) systems at 0.1 MPa and 298.15 K. Numbers in parentheses indicate the number of atoms of the 1-alkanol.